\documentclass{svmult}
\pdfoutput=1 % flag that we are using pdflatex not latex

\usepackage{makeidx}
\usepackage{graphicx}
\usepackage{cite}
\usepackage{hyperref}
\usepackage[bottom]{footmisc}% places footnotes at page bottom

\def\descitem#1{\medskip\item[\textbf{#1}]}

\makeindex             % used for the subject index
\begin{document}

\title*{Growth and Decay in\\ Life-Like Cellular Automata}
\titlerunning{Growth and Decay}

\author{David Eppstein}
\institute{Computer Science Department\\
University of California, Irvine\\
Irvine, CA 92697-3435, USA\\
\texttt{eppstein@uci.edu}}
\maketitle
\section{Introduction}

Since the study of life began, many have asked: is it unique in the universe, or are there other interesting forms of life elsewhere? Before we can answer that question, we should ask others: What makes life special? If we happen across another system with life-like behavior, how would we be able to recognize it? We are speaking, of course, of the mathematical systems of cellular automata, of the fascinating patterns that have been discovered and engineered in Conway's Game of Life~\cite{BerConGuy-04,Gar-SA-70}, and of the possible existence of other cellular automaton rules with equally complex behavior to that of Life.

In an influential early paper~\cite{Wol-PD-84}, Stephen Wolfram\index{Wolfram, Stephen}
proposed an answer to this question of what makes Life special. He categorized cellular automaton rules into four types, according to their behavior when started with random initial conditions:
\begin{itemize}
\item In Class I automata, all cells eventually become the same.
\item In Class II, the field eventually degenerates into scattered stable or oscillating patterns.
\item In Class III, the chaos of the initial random pattern persists indefinitely.
\item Class IV contains the remaining rules, in which patterns such as those in Life exhibit complex behavior.
\end{itemize}
Wolfram initially intended this classification for one-dimensional cellular automata,\index{one-dimensional cellular automaton} but in a later paper~\cite{WolPac-JSP-85} he and Packard extended it to the same set of two-dimensional automata that we consider here;  Adamatzky et al.~\cite{AdaMarMor-IJBC-06}\index{Mart{\'\i}nez , Genaro Ju{\'a}rez}\index{Adamatzky, Andrew}\index{Mora, Juan Carlos Seck Tuoh} performed a more detailed classification of some of these automata based on the same principle of studying their behavior on random initial conditions.

However, Wolfram's classification is problematic in more than one way.  Although there is some evidence for phase transitions\index{phase transition} between regions of rule space where one class is more frequent than others~\cite{ChaMan-PD-90,LiPacLan-PD-90,WooLan-PD-90,LafBos-RAAL-05}, the classification depends strongly on the specific behavior of an individual rule, so that one cannot use it to predict which rules are likely to have interesting behavior, but only to describe that behavior after having already observed it. The boundary between classes is less clear-cut and more subjective than one would like, and may for some automata be impossible to decide~\cite{McI-PD-90,CulYu-CS-88}.\index{undecidability} Some automata may have a constant probability of behaving in more than one of these ways~\cite{BalShe-TCS-00}, making them difficult to classify. It is not obvious, even, which class Life properly belongs to, and it is conventionally classified as Class IV less because the description of that class best fits our observations of Life and more because Life is the archetypical rule whose behavior Class IV was intended to capture. When the boundary is clear, it is not where we might like it to be: for instance, the rule B35/S236 (discussed below) appears to be in Class III, but can support many complex patterns similar to those in Life. Wolfram's classification defines the interesting rules negatively rather than positively: they are the rules where some known type of uninteresting behavior doesn't happen. It is predicated on the assumption that interesting behavior should emerge from uninformative initial conditions, but many of the most interesting patterns in Life could not have been found in this way.

We propose a four-way classification of two-dimensional semi-totalistic cellular automata that is different than Wolfram's, based on two questions with yes-or-no answers: do there exist patterns that eventually escape any finite bounding box placed around them? And do there exist patterns that die out completely? If both of these conditions are true, then a cellular automaton rule is likely to support spaceships,\index{spaceship} small patterns that move and that form the building blocks of many of the more complex patterns that are known for Life. If one or both of these conditions is not true, then there may still be phenomena of interest supported by the given cellular automaton rule, but we will have to look harder for them. Although our classification is very crude, we argue that it is more objective than Wolfram's (due to the greater ease of determining a rigorous answer to these questions), more predictive (as we can classify large groups of rules without observing them individually), and more accurate in focusing attention on rules likely to support patterns with complex behavior. We support these assertions by surveying a number of known cellular automaton rules.

\section{Life-like rules}

The space of possible cellular automaton rules is infinite and highly varied. One may define cellular automata on grids of high dimensions or on neighborhood structures more general than grids. The set of neighbors of a cell may be only those other cells nearest to it in the grid or may fall within a neighborhood of larger than unit radius. The state of a cell may depend only on the states of neighboring cells in the previous time step, or it may depend on the states of neighbors over several previous time steps. The number of states of each cell may be any finite number or even a continuously variable value, and researchers have considered update rules that are asynchronous,\index{asynchronous cellular automaton} randomized,\index{nondeterministic cellular automaton} or quantum mechanical.\index{quantum cellular automaton}

In order to impose some order on this vast wilderness, we restrict our attention to a more circumscribed set of rules that are very similar in structure to Conway's Game of Life. Specifically, we consider semi-totalistic\index{semi-totalistic rule} (or outer totalistic)\index{outer totalistic rule} binary cellular automata on a two-dimensional Moore neighborhood.\index{Moore neighborhood} These are the cellular automata in which:

\begin{itemize}
\item The cells of the automaton form a two-dimensional square lattice.

\item The neighbors of each cell are the eight lattice squares that are orthogonally or diagonally adjacent to it.

\item Each cell may be in one of two states, alive or dead.

\item All cells are updated simultaneously in a sequence of time steps.

\item In time step $i$, the state of any given cell is a function of the state of the same cell in time step $i-1$ and of the number of live neighbors it had in time step $i-1$.
\end{itemize}

In a cellular automaton of this type, a single cell may do one of four things within a single time step: If it was dead but becomes alive, we say that it is \emph{born}.\index{birth} If it was alive and remains alive, we say that it \emph{survives}.\index{survival} If it was alive and becomes dead, we say that it \emph{dies}.\index{death} And if it was dead and remains dead, we say that it is \emph{quiescent}.\index{quiescence}

We follow a standard convention for naming these cellular automata in which the update rule of the automaton is represented by a \emph{rule string}\index{rule string}, a sequence of characters in the form ``B\textit{xxx}/S\textit{yyy}''. The \textit{xxx} part of the rule string is a subset of the digits from 0 to 8, representing numbers of neighbors such that a dead cell with that many neighbors would become alive in the next time step, causing a birth event: the B stands for birth. The \textit{yyy} part of the rule string is another subset of digits, representing numbers of neighbors such that a live cell with that many neighbors would remain alive in the next time step, causing a survival event: the S stands for survival. For instance, Conway's Game of Life itself is represented by the rule string ``B3/S23'': a dead cell with three live neighbors leads to a birth event, and a live cell with two or three live neighbors leads to a survival event. All other combinations of cell state and number of neighbors lead to death or quiescence; we do not need to list these combinations separately as they can be inferred from the birth and survival parts of the rule string.

Each digit from 0 to 8 may be present or absent in the birth part of the rule string, and may independently be present or absent in the survival part. Therefore, there are $2^{18}$ different Life-like rules, too many to study in detail individually: we need a roadmap to help guide us to the interesting rules.

\section{Natural evolution or intelligent design?}

Wolfram's classification takes the point of view of statistical mechanics,\index{statistical mechanics} a field that uses probability to study large systems of interacting objects such as the molecules in an ideal gas.\index{ideal gas} Wolfram posits that a typical initial state\index{initial state} of the automaton has a 50/50 chance that any given cell is alive or dead. independently of all other cells; he then asks how the automaton behaves when started from this typical state.

From the mathematical point of view, random initial conditions are completely general: any finite configuration of an automaton, such as a glider gun centered within an otherwise quiescent  $10^6\times 10^6$ square of empty space, occurs infinitely often within a random initial state, despite the tiny probability ($2^{-10^{12}}$) that this pattern occurs at any particular location.
However, from the practical point of view, the cellular automaton patterns that can be seen to occur randomly and the patterns that can be constructed by human engineering are very different from each other. The patterns arising from random fields in Life, as seen in practice, consist overwhelmingly of small still lifes\index{still life} and small period-two oscillators\index{oscillator}, with higher-period oscillators such as the period-3 pulsar,\index{pulsar} period-15 pentadecathlon,\index{pentadecathlon} and period-8 figure-8\index{figure-8} arising much less frequently (once per 20,000, 1.6 million, or 33 million distinct patterns generated in this way, respectively) in experiments performed by Achim Flammenkamp\index{Flammenkamp, Achim}~\cite{Fla-04}. In contrast, many of the most interesting patterns in Life were formed by human engineering, and may combine hundreds of individual patterns that themselves have much higher complexity than the patterns that could ever be seen in random experiments by any human observer. If we seek rules that can support similar patterns, we should use a classification scheme that classifies those rules similarly to Life, regardless of whether they behave similarly on random initial conditions.
Consider, for instance, the following Life patterns:

\begin{figure}[t]
\centering\includegraphics[width=4in]{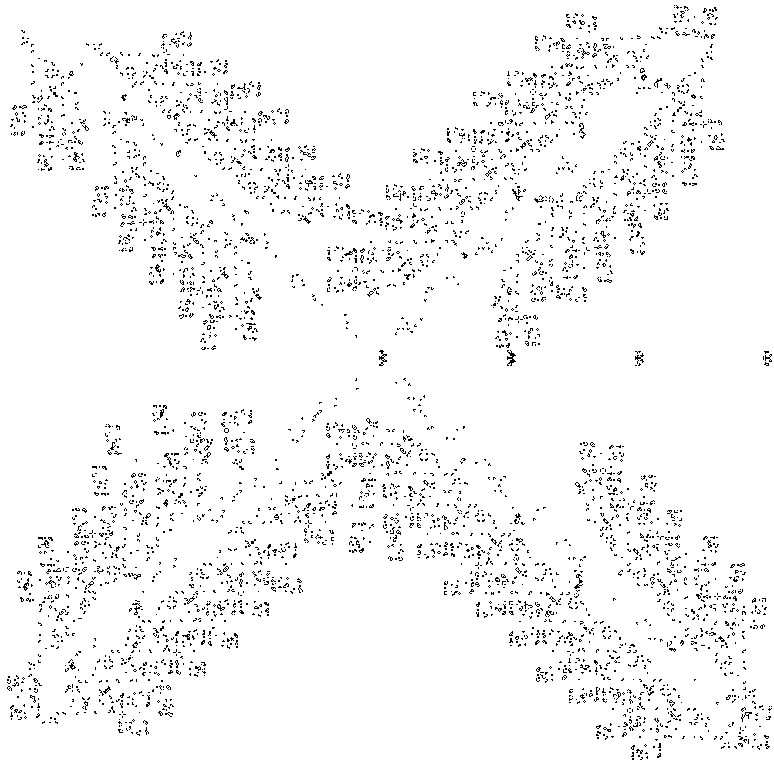}
\caption{Greene's period-416 $2c/5$ spaceship gun.}
\label{fig:2c5p416}
\end{figure}

\begin{itemize}
\item The pattern library included with the Golly\index{Golly} life simulation software\cite{Golly} includes a period-416 $2c/5$ gun, designed in 2003 by Dave Greene\index{Greene, Dave} based on a reaction due to Noam Elkies\index{Elkies, Noam} (Fig.~\ref{fig:2c5p416}). It fits within a square of 1 million cells; every 416 time steps it produces a spaceship that moves at speed $2c/5$. It consists of 64 period-416 glider guns; each of these glider guns in turn is formed by five small still life patterns, four small low-period oscillators, and a larger high-period pattern that is contained and controlled by its interactions with the still lifes and oscillators. 60 separate patterns, each of which is again formed from a cluster of still lifes and low-period oscillators, convert the gliders from these guns into Herschel patterns\index{Herschel pattern} (highly reactive patterns formed from seven live cells) and then back into gliders in order to insert them into position within four diagonal glider streams that converge in the center of the pattern, where 64 gliders repeatedly react with each other to build up complex still life patterns and then trigger them to become the $2c/5$ spaceships that are emitted by the gun. Overall, the pattern has over 26,000 live cells.
\item Brice Due's\index{Due, Brice} ``Outer Totalistic Cellular Automata Meta-Pixel''~\cite{Due-06} allows any Life-like cellular automaton to be simulated within Life, by representing each cell of the other automaton as a ``meta-pixel''\index{meta-pixel} of approximately 4 million Life cells; it takes 35328 steps of Life for the meta-pixel to simulate a single step of the other automaton. The interior of each meta-pixel is either quiescent (when it represents a dead cell of the simulated automaton) or covered by lightweight spaceships\index{lightweight spaceship} when it represents a live cell. The boundary area of each meta-pixel is filled with period-46 oscillators and other components which control whether the spaceships are generated or not, generate the spaceships, and communicate via gliders to the corresponding components of adjacent meta-pixels in order to compute the next state at each time step of the automaton. Many copies of the meta-pixel pattern must be combined to simulate any nontrivial pattern of another automaton.
\item The Caterpillar~\cite{Niv-05}\index{caterpillar} is a large spaceship that moves at speed 17c/45, constructed in 2004 by David Bell,\index{Bell, David} Gabriel Nivasch,\index{Nivasch, Gabriel} and Jason Summers\index{Summers, Jason}. The pattern repeats its configuration every 270 steps, advancing by 102 cells in an axis-parallel direction within that time period. It is based on a reaction in which the \emph{$\pi$ heptomino} pattern\index{pi heptomino@$\pi$ heptomino} moves at this speed along a track of properly spaced blinkers without destroying the track. If two $\pi$ heptominos move in tandem along parallel tracks, they can be made to emit gliders; interactions between the gliders emitted by multiple heptominos sharing a system of tracks can be used to tear down the tracks behind the heptominos and to send streams of faster spaceships toward the front of the moving system of heptominos. These spaceships interact with additional gliders to extend the tracks on which the heptominos follow. The whole pattern uses large numbers of these components; it fits within a bounding box with an area of 1.4 billion cells, of which approximately 12 million are alive at any time.
\end{itemize}

Beyond the practical unlikelihood of discovering patterns such as these from random initial conditions, there is another reason that using random initial states to classify automata is problematic: with this assumption, it is very difficult to say anything that can be backed up by rigorous mathematics. For the Game of Life, what we know rigorously is limited to random states in which the probability of a cell being live is a number $\epsilon$ that is very close to zero, on time and distance scales that are bounded by polynomial functions of $1/\epsilon$~\cite{Got-NCCA-03}. For random states with greater numbers of live cells, or over a greater number of time steps, much remains unproven. For instance, in a random Life universe, does every cell eventually becomes periodic with probability one? If so, Life should probably be classified as Class II rather than Class IV. There exist finite initial configurations for Life that do not ever become periodic, but we do not know whether it is possible for nonperiodic configurations to survive when surrounded by the ashes of a random starting state.

It is perhaps worth emphasizing that this dichotomy between nature and design in Conway's Life does not tell us much about the proper study of nature itself. The physical universe is enormously larger and has been running on far longer time scales than any cellular automaton simulation, and we have plenty of evidence in front of and behind our eyes that complex systems can evolve from simpler ones. For all we know, it may be possible that a Life simulation over similarly large time and space scales, from an initially random state, could eventually develop something recognizable as an ecology\index{ecology}~\cite{BerConGuy-04} rather than becoming periodic everywhere. In this respect, the frequently used term ``evolution''\index{evolution} for the behavior of a cellular automaton pattern over a sequence of time steps is unfortunate, because at the scales we can observe this behavior is much better modeled as physics than as biology.

\section{Growth}

Because our primary interest is in engineered patterns rather than random fields, we restrict our attention to patterns that have a finite number of live cells. This eliminates the possibility of rules with births on zero live neighbors, because in such rules any finite pattern would immediately become infinite. Without B0, we need consider only $2^{17}$ possible rules, half as many as before.

A \emph{bounding box}\index{bounding box} for a pattern is a rectangle with axis-parallel sides that contains all the live cells of the pattern; the \emph{minimum bounding box} is the smallest possible such rectangle, but we also allow larger rectangles to count as bounding boxes. Every pattern with finitely many live cells can be placed within a bounding box.

We define a cellular automaton rule to be \emph{fertile}\index{fertile} if it has a finite pattern that eventually escapes any of its bounding boxes. That is, the rule is fertile if there exists a \emph{growth pattern}\index{growth pattern} $P$ such that, for every bounding box $B\supset P$, after some number of time steps starting from $P$ there will be a live cell outside $B$. The growth patterns in Life include many types of pattern that Life enthusiasts have found interesting, including gliders and spaceships, puffer trains,\index{puffer train} guns, rakes,\index{rake} breeders,\index{breeder} and other spacefillers.\index{spacefiller}

If a rule is not fertile, then every finite pattern must eventually become periodic: for every pattern $P$ there is a bounding box $B$ that it cannot escape, and it can only progress through $2^{|B|}$ possible states before returning to a state that it has already been in. On the other hand, if every finite pattern eventually becomes periodic, then there is no pattern that can escape all of its bounding boxes. That is, instead of defining fertility in terms of escape from bounding boxes, it would be equivalent to define an \emph{infertile} rule as one in which every finite pattern eventually becomes periodic. However, the definition in terms of bounding boxes and growth patterns is more convenient when attempting to determine which rules are fertile and which are infertile.

\begin{figure}[t]
\centering\includegraphics{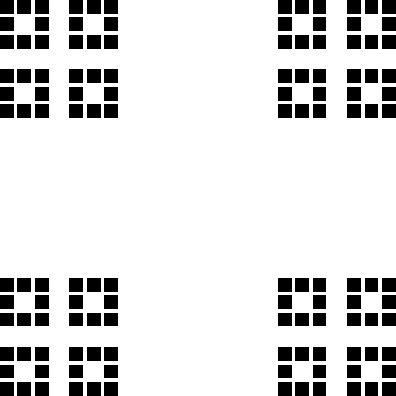}
\qquad\qquad\includegraphics{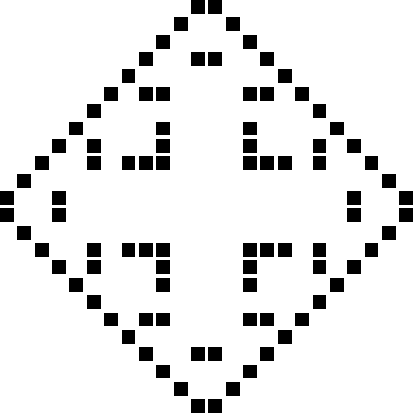}
\caption{Growth patterns in B1 and B2 rules. Left: 11 steps after starting from a single live cell in B1/S. As is true for all steps of this pattern in all B1 rules, the four corner cells of the minimal bounding box are alive. Right: 11 steps after starting from a $2\times 2$ block of live cells in B2/S. As is true for all steps of this pattern in all B2 rules, in each edge of the minimal bounding box only the two middle cells are alive.}
\label{fig:growth}
\end{figure}

Some simple case analysis allows us to determine whether a rule is fertile for most of the $2^{17}$ rules that support finite patterns:
\begin{itemize}
\item If a rule includes B1, it is fertile. In this case, the pattern consisting of a single live cell is a growth pattern. After $k$ steps starting from this pattern, the minimum bounding box will contain $(2k+1)\times(2k+1)$ cells, and will have a single live cell at each of its four corners (Fig.~\ref{fig:growth}, left).
\item If a rule includes B2, it is fertile. If the rule also includes B1, this follows from the previous case; otherwise, the pattern consisting of a $2\times 2$ block of live cells is a growth pattern. After $k$ steps starting from this pattern, the minimum bounding box will contain $(2k+2)\times (2k+2)$ live cells, and will have two adjacent live cells at the center of each of its edges; the remaining cells on each edge of the bounding box will be non-live, leading the same pattern to propagate outwards by one more unit in the following time step (Fig.~\ref{fig:growth}, right).
\item If a rule does not include B1, B2, or B3, it is not fertile. The dead cells outside of a bounding box $B$ of any pattern can have at most three live neighbors, and therefore can never become live themselves.
\end{itemize}
The remaining cases are those with rule strings that begin ``B3\dots'', as Life's rule string B3/S23 does.  For these rules, often the simplest way to show that they are fertile is to exhibit a growth pattern such as Life's glider. We have used our search software~\cite{Epp-MSRI-02}, together with searches using small random seeds, to search for spaceships in these B3 rules; so far, we have found that 10736 out of the 16384 possible B3 rules have spaceships and therefore are fertile~\cite{Fano}.

\begin{figure}[t]
\centering\includegraphics[scale=0.35]{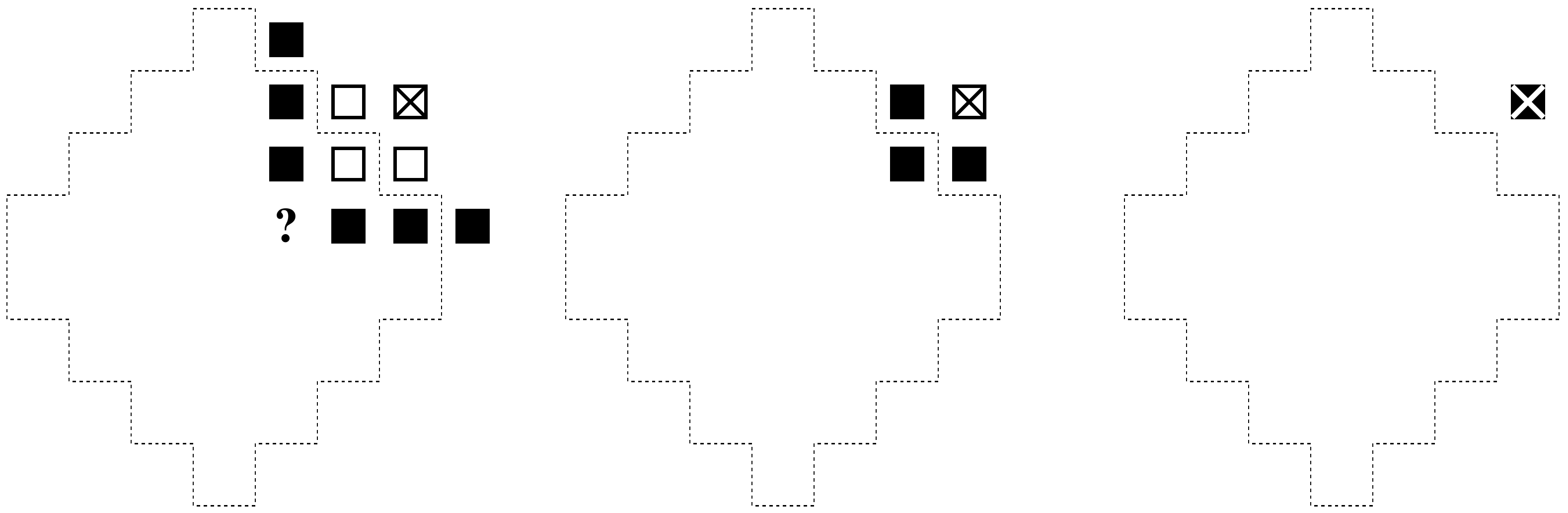}
\caption{Infertility in rules such as B36/S678 without births on 2, 4, or 5 neighbors and without survival on fewer than six neighbors. In order for a point $x$ that has only one neighbor in an initial bounding diamond to become live in step~$i$ (right), three neighbors of $x$ closer to the bounding diamond must be live in step~$i-1$ (center), but there is no way to configure the cells in step $i-2$ to cause these three neighbors to be born (left).}
\label{fig:infertility}
\end{figure}

In some of the remaining cases, it is possible to prove using a more detailed analysis that a rule is not fertile. For instance, suppose that a B3 rule does not allow births with 1, 2, 4, or 5 live neighbors, nor does it allow survivals with five or fewer live neighbors. In this case, no growth pattern can exist. For, let $P$ be any pattern, and let $D$ be a bounding diamond of $P$ (that is, a shape containing it bounded by lines of slope $\pm 1$) We prove by contradiction that no cell that has at most one neighbor in $D$ can ever become live in any future state of $P$. For, otherwise, suppose that $x$ is the first such cell to become live, at time step~$i$; in Fig.~\ref{fig:infertility}, right, $x$ is the dark marked cell, and $D$ is the dotted polygon. Then in order to become born, $x$ must have three live neighbors at time step~$i-1$: two neighbors $y$ and $z$ of $x$ are live and outside $D$, and a third neighbor $w$ is live and inside $D$.  These four cells $w$, $x$, $y$, and $z$ form a $2\times 2$ block, as that is the only way to arrange $y$ and $z$ while allowing them to have more than one neighbor in $D$---see Fig.~\ref{fig:infertility}, center. Before step $i$, every cell outside $D$ has too few live neighbors to survive, so $y$ and $z$ must have been newly born at step $i-1$, and were not live at step $i-2$. At step $i-2$, $w$ has five or fewer live neighbors, for its three neighbors $x$, $y$, and $z$ are not live; therefore, if it were live at that step it would have died, contradicting its living state in step $i-1$. Therefore, at step $i-1$, $w$ is also newly born. However, it is not possible for $w$, $y$, and $z$ to all be newly born in the same step: the three live neighbors needed for $y$ to be newly born and the three live neighbors needed for $z$ to be newly born (Fig.~\ref{fig:infertility}, left) cause $w$ to have four or five live neighbors and to remain quiescent. This contradiction implies that $P$ cannot escape an expanded bounded diamond surrounding $D$ and therefore that the rule is not fertile. There are 64  B3 rules without any of B1245/S012345, all of which can be proven to be infertile using the above analysis.

There remain only 5584 rules for which we neither have a known growth pattern nor a proof of infertility. Therefore, this analysis allows us to determine in 96\% of the cases whether a rule is fertile or infertile. Additionally, this analysis ignores the possibility of growth patterns other than spaceships, such as the ladders that are known in Life Without Death\index{Life Without Death} (B3/S012345678),\index{B3/S012345678} Maze\index{Maze} (B3/S12345),\index{B3/S12345} and related rules.

A related attempt to classify cellular automata by their growth properties was made by Gravner\index{Gravner, Janko}~\cite{Gra-NCCA-03}. Gravner considered initial patterns formed by setting the cells within an $n\times n$ bounding box to be alive or dead randomly, and by setting the cells outside the bounding box to be dead. A pattern of this type exhibits \emph{quadratic growth} (or, in Gravner's terminology, \emph{linear expansion})  if, after $t$ steps of the automaton, it has $\Omega(t^2)$ live cells, so that it eventually grows to fill a large fraction of the plane.
Gravner defined a cellular automaton rule to be \emph{expansive} if, with probability approaching one in the limit as $n$ goes to infinity, random patterns with $n\times n$ bounding boxes exhibit linear expansion. However, being able to cover the plane is a much stricter requirement on a pattern than being able to escape a bounding box, so it is often difficult to determine whether a pattern exhibits linear expansion, and even more difficult to determine whether a rule is expansive. More, this classification shares with Wolfram's classification the property that it is based on random initial conditions, so (as we have argued above) it does not address well the ability of a cellular automaton to support non-random structures.

\section{Decay}

In order for its patterns to exhibit the complex behavior that they do, it is important in Life that some patterns shrink as well as that others grow. In the extreme, some patterns may eventually lead to a state in which every cell is dead and quiescent; in the terminology of \emph{Winning Ways}\index{Winning Ways}~\cite{BerConGuy-04}, a pattern of this type is said to \emph{fade}. For instance, the proof that determining the eventual fate of a Life pattern is undecidable\index{undecidability} depends on patterns that fade: it is undecidable to determine, for a given Life pattern, whether it fades or whether it has some living cells in every future state~\cite{BerConGuy-04}.

\begin{figure}[t]
\centering\includegraphics[width=4in]{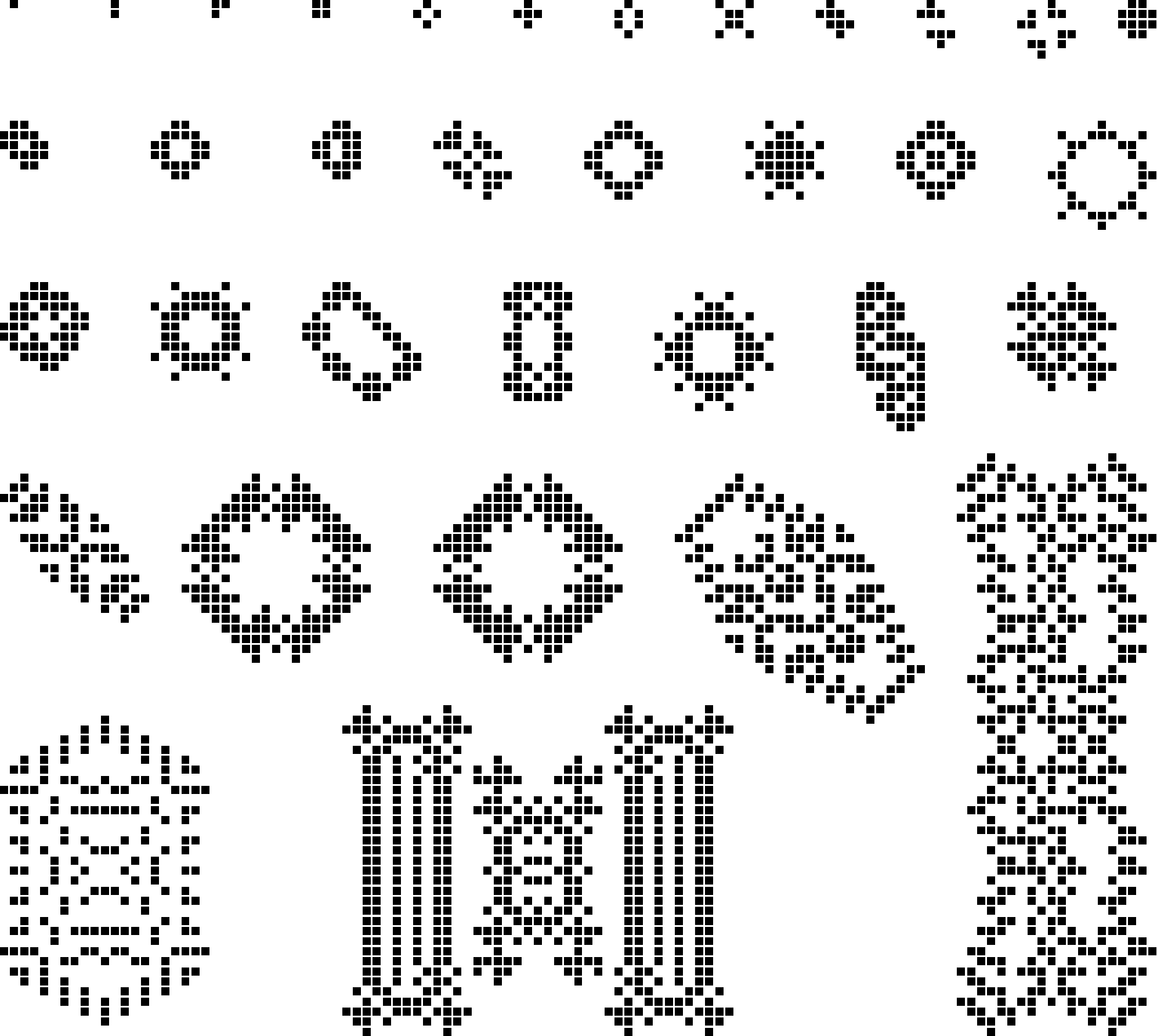}
\caption{Patterns that fade in any rule that includes none of the following birth and survival possibilities.
Row~1: B1/S0, B12/S1, B123/S2, B12/S3, B124/S2, B124/S2, B13/S34,
B125/S2, B13/S14, B13/S35, B135/S23, B134/S2, B123/S47.
Row~2: B1238/S46, B1235/S45, B1237/S456, B135/S245,
B1234/S45, B134/S1578, B1236/S45, B13/S13.
Row~3: B12347/S46, B134/S156, B1235/S46, B12346/S46, B1345/S157, B12367/S46, B1357/S256.
Row~4: B1356/S246, B135/S2567, B135/S2578, B1357/S246, B1356/S24.
Row~5: B135/S12, B1356/S25.}
\label{fig:diehard}
\end{figure}

If a rule supports a pattern $P$ with a finite number of live cells, such that the state following $P$ has no live cells, we say that the rule is \emph{mortal}, and otherwise that it is \emph{immortal}. It is equivalent to ask whether the rule has a finite pattern that fades, because if pattern $Q$ fades, the pattern $P$ formed from $Q$ on the penultimate step before it fades has the desired property that the state following $P$ is empty. However, requiring that $P$ fade in the next step rather than at some future time simplifies our analysis.

As Dean Hickerson\index{Hickerson, Dean} observed, a pattern fades in rule $r$ if and only if the same pattern forms a still life\index{still life} in the rule $\bar r$ with the same birth conditions and complementary survival conditions. A 2001 analysis by Hickerson, aided by Matthew Cook,\index{Cook, Matthew} Jason Summers,\index{Summers, Jason} and the author, determined the existence or nonexistence of a still life in most of the $2^{17}$ possible rules without B0, and the same analysis can also be used to determine which of these rules are mortal. In particular, in the rules that are known to be mortal, one of the patterns shown in Figure~\ref{fig:diehard} must fade.
We provide some flavor of the analysis of immortal rules, although not the complete analysis, below.

\begin{itemize}
\item If a rule allows births with exactly one live neighbor, then it is immortal. For, if $(x,y)$ are the Cartesian coordinates of a live cell in a pattern that maximizes $y$ among all live cells in a given finite pattern, and that maximizes $x$ among all live cells with the same $y$-coordinate, then in the next step cell $(x+1,y+1)$ will also be live---see Fig.~\ref{fig:immortal4}, far left.
\item If a rule causes all live cells with fewer than five live neighbors to survive, then it is immortal. For, if $(x,y)$ is a live cell that maximizes $y$ among all live cells in a given finite pattern, and that maximizes $x$ among all live cells with the same $y$-coordinate, then $(x,y)$ has fewer than five live neighbors and must survive into the next step---see Fig.~\ref{fig:immortal4}, far left again.
\item If a rule allows births with both two and three live neighbors, and survival with zero live neighbors, then it is immortal. In this case the analysis showing immortality is somewhat more intricate. Suppose that, in a given pattern $P$ with finitely many live cells, $(x,y)$ is the live cell maximizing $x+y$, and that among all live cells with that value of $x+y$ it is the one maximizing $y$. Further, suppose for a contradiction that the next state from $P$ according to the given rule has all cells dead. Then cell $(x-1,y)$ must be dead in $P$, for otherwise there would be a birth at $(x,y+1)$. Cells $(x,y-1)$ and $(x+1,y-1)$ must also be dead, for otherwise there would be a birth at $(x+1,y)$. This eliminates all possible live neighbors of $(x,y)$ except for $(x-1,y-1)$, which must be live in order for $(x,y)$ to die at the next step. Additionally, $(x-2,y)$ and $(x-2,y+1)$ must be dead, for otherwise there would be a birth at $(x-1,y+1)$. All of these conditions together imply that cell $(x-1,y)$ has either two or three live neighbors: $(x,y)$ and $(x-1,y-1)$ are live, and all other neighbors with the possible exception of $(x-2,y-1)$ are dead. But then, $(x-1,y)$ would have a birth in the next state, contradicting our assumption. This case is shown in Fig.~\ref{fig:immortal4}, center left.

\begin{figure}[t]
\centering\includegraphics[scale=0.35]{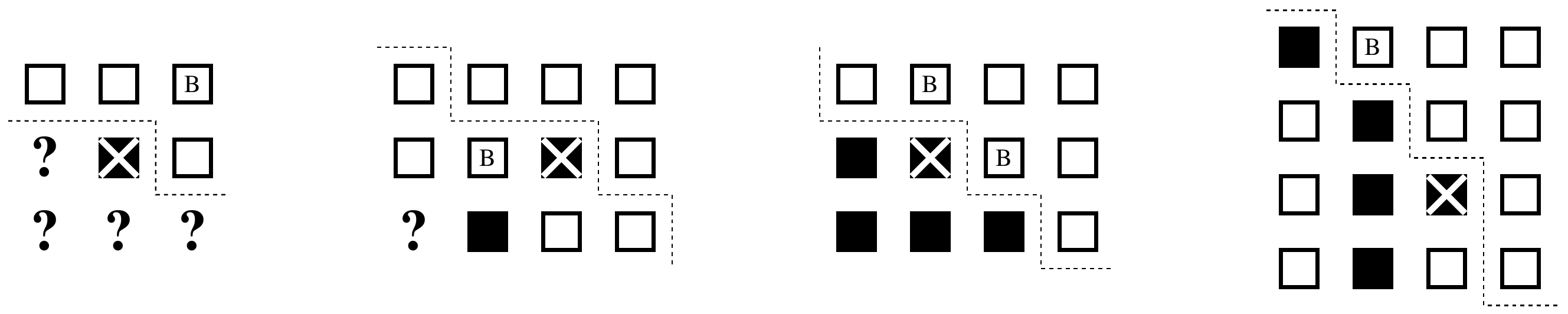}
\caption{Case analysis for immortality of certain rules. In each case the dark squares show live cells, the open boxes show dead cells, and the dark square marked with a white X is the cell $(x,y)$ selected at the start of the case. Due to the way $(x,y)$ was selected, all cells above and to the right of the dashed line must be dead. An open box with the letter B in it marks a location of cell birth. Far left: rules with B1 are immortal due to a birth at $(x+1,y+1)$, while rules with S01234 are immortal because no matter how the cells marked by question marks are set, cell $(x,y)$ survives. Center left: Rules with B23/S0 lead to a birth at the marked square. Center right: Rules with B2/S0123 or B3/S0123 lead to a birth at one of the two marked squares. Far right: Rules with B2/S01245 lead to a birth at the marked square.}
\label{fig:immortal4}
\end{figure}

\item If a rule allows births with either two or three live neighbors, and it causes any live cell with fewer than four live neighbors to survive, then is immortal. For, if $(x,y)$ is a live cell that maximizes $x+y$ among all live cells in a given finite pattern, and that maximizes $y$ among live cells with the same value of $x+y$, then $(x,y)$ can have at most four live neighbors, at $(x-1,y)$, $(x-1,y-1)$, $(x,y-1)$, and $(x+1,y-1)$. If fewer than four of these neighbors are live then $(x,y)$ survives, and if all four are live then there is a birth at $(x,y+1)$ (for rules with B2) or $(x+1,y)$ (for rules with B3)---see Fig.~\ref{fig:immortal4}, center right.
\item If a rule includes all of B2/S01245, then it is immortal. For, let $(x,y)$ be a live cell that maximizes $x+y$ among all live cells in a given finite pattern $P$, and that maximizes $x$ among all live cells with the same value of $2x+y$. Then, if $P$ is to fade on the next step, $(x,y-1)$ must be dead, for otherwise there would be a birth at $(x+1,y)$. Thus $(x,y)$ has at most three live neighbors, and if it is to die on the next step it must have exactly three, at each of the remaining neighboring locations $(x-1,y+1)$, $(x-1,y)$, and $(x-1,y-1)$. This also provides $(x-1,y)$ with three live neighbors; if it is to die, its neighbor count must be either three or six, and if it had six live neighbors then $(x-1,y+1)$ would have four or five and would survive. We can therefore conclude that the three locations $(x-2,y+1)$, $(x-2,y)$, and $(x-2,y-1)$ are all dead. In order for $(x-1,y+1)$ to avoid surviving, it needs a third live neighbor at $(x-2,y+2)$; but then, cell $(x-1,y+2)$ would have two live neighbors and would lead to a birth in the next step---see Fig.~\ref{fig:immortal4}, far right. We conclude that $P$ cannot fade and therefore that the rule is immortal.

\item If a rule includes all of B345/S013, then it is immortal. For, let $(x,y)$ be a live cell that maximizes $x+y$ among all live cells in a given finite pattern, and that maximizes $y$ among live cells with the same value of $x+y$. Then, in order for $(x,y)$ to die in the next step, it must have two or four live neighbors among the four cells $(x-1,y)$, $(x-1,y-1)$, $(x,y-1)$, and $(x+1,y-1)$. However, four live neighbors would lead to a birth at $(x+1,y)$, so the only possibility for the entire pattern to fade in the next step is to have two live neighbors. We may now consider sub-cases, according to how those live neighbors of $(x,y)$ are arranged, to show that in each case the pattern has a birth or survival into the next step:
\begin{itemize}
\item If $(x,y-1)$ is not live, then it has as live neighbors $(x,y)$ and the other two live neighbors of $(x,y)$. In order to prevent a birth from occurring at $(x,y-1)$ in the next step of the pattern, the three cells $(x-1,y-2)$, $(x,y-2)$, and $(x+1,y-2)$ must also all be live. But with these choices fixed, the cell $(x+1,y-1)$ has either three or four live neighbors, and would, if dead, be the location of a birth in the next step. Thus, $(x+1,y-1)$ must be live, and in order for it to die in the next step $(x+2,y-2)$ must also be live---see Fig.~\ref{fig:b345s013}, far left.
\item If $(x,y-1)$ is live, and the other live neighbor of $(x,y)$ is at $(x+1,y-1)$, then there is a birth at $(x+1,y)$---see Fig.~\ref{fig:b345s013}, center left.
\item If $(x,y-1)$ is live, and the other live neighbor of $(x,y)$ is at $(x-1,y)$, then $(x-2,y)$ must be dead, for otherwise there would be a birth in the next step at $(x-1,y+1)$.  $(x,y-2)$ and $(x+1,y-2)$ must be dead, for otherwise there would be a birth at $(x+1,y-1)$. And $(x-1,y-2)$ must be dead, for otherwise the cell at $(x,y-1)$ would survive. But these choices together imply that $(x-1,y-1)$ has from three to five live neighbors, leading to a birth there in the next step---see Fig.~\ref{fig:b345s013}, center right.
\item If $(x,y-1)$ is live, and the other live neighbor of $(x,y)$ is at $(x-1,y-1)$, then the three cells $(x-2,y-1)$, $(x-2,y)$, and $(x-2,y+1)$ must all be live to prevent a birth at $(x-1,y)$. But then there would be a birth at $(x-1,y+1)$---see Fig.~\ref{fig:b345s013}, far right.
\item A similar analysis shows that the rules
B/S0123567,
B2/02345,\\
B2/S023467,
B24/S0234,
B24/S01345,
B245/S01356,
B246/S013467,
B256/S023468,
B257/S023468,
B2456/S013,
B2457/S0135,\\
B2467/S013468,
B24678/S013478,
B456/S012367,
B4568/S012357,\\
B45678/S01237,
B4578/S012357, and
B5678/S012357
are all immortal, as are any rules that include births or survivals on a superset of the numbers of neighbors of these rules.
\end{itemize}
\end{itemize}

\begin{figure}[t]
\centering\includegraphics[scale=0.35]{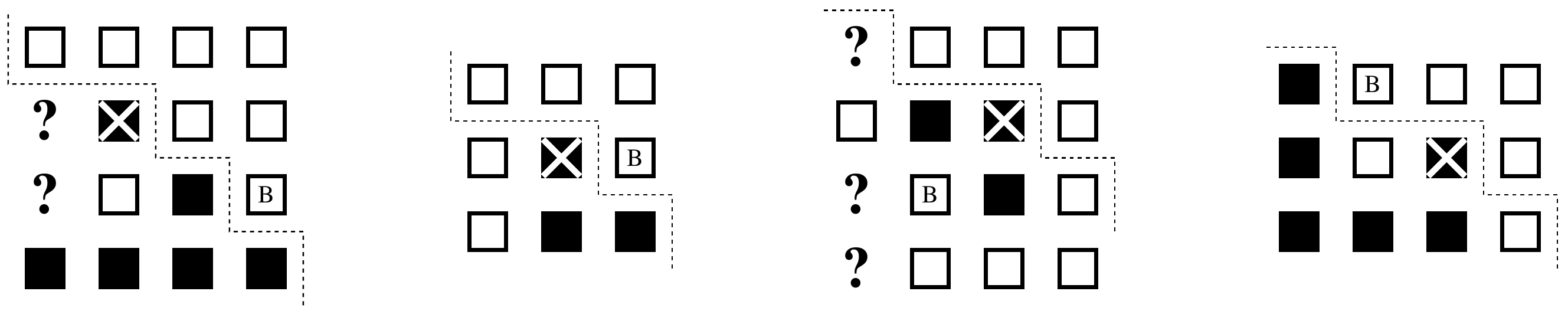}
\caption{Case analysis for immortality of rules with B345/S013. Far left: $(x,y-1)$ is not live. Center left: $(x,y-1)$ and $(x+1,y-1)$ are live.
Center right: $(x,y-1)$ and $(x-1,y)$ are live. Far right: $(x,y-1)$ and $(x-1,y-1)$ are live.}
\label{fig:b345s013}
\end{figure}

These cases together show that, of the 131072 possible life-like rules without B0, 77563 of them are immortal and 53214 of them are mortal. There remain 295 rules that are not classified by this analysis, so we may determine whether a rule is mortal or not in approximately 99.8\% of the cases. In particular this classification covers all rules that include B3.

\begin{figure}[t]
\centering\includegraphics{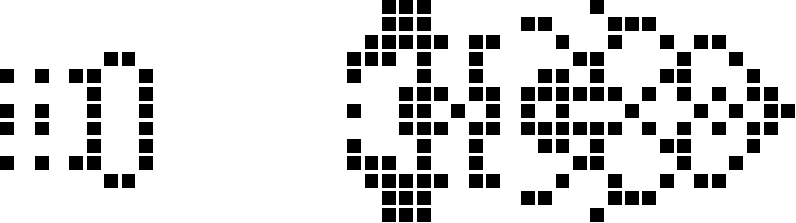}
\caption{A $c/4$ orthogonal spaceship (left) and a $c/3$ orthogonal spaceship (right) in the immortal rule B3456/S013.}
\label{fig:b3456s013}
\end{figure}

It does not follow from the definition of mortality that an immortal rule can have no spaceships. For instance, the immortal rule B3456/S013\index{B3456/S013} has spaceships, shown in Fig.~\ref{fig:b3456s013}. 
However, in many of the cases for which we can prove that a rule is immortal, the proof shows something stronger, that the minimal bounding box or minimal bounding diamond of a pattern can never shrink. When this is true, it is impossible for a spaceship to exist.

\section{The Life-like menagerie}

\begin{figure}[p]
\centering\includegraphics[width=3in]{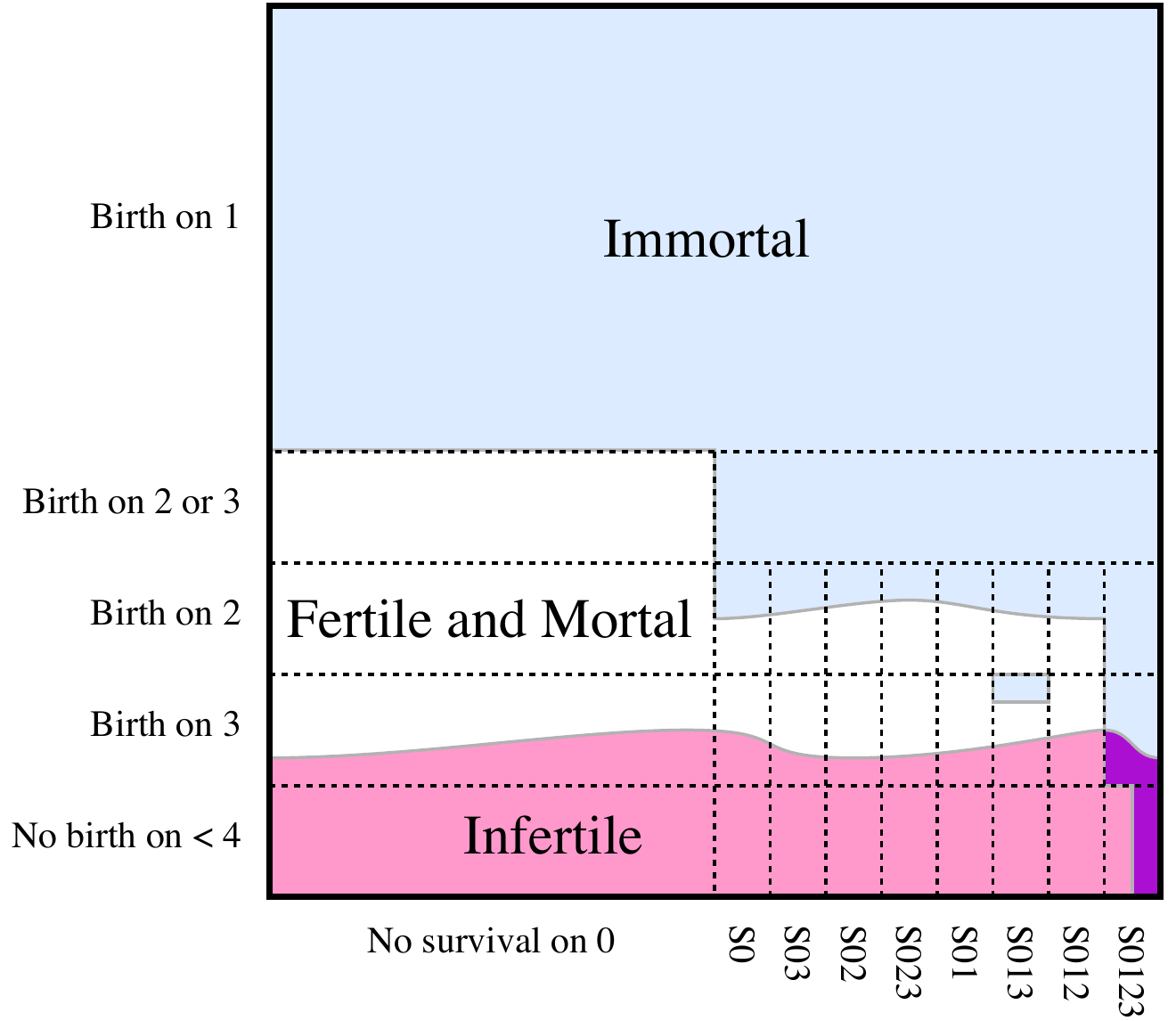}
\caption{A map of the possible life-like rules, depicting regions of fertile, infertile, mortal, and immortal rules. The area of a region represents approximately the number of different rules within it.}
\label{fig:map}
\end{figure}
\begin{figure}[p]
\centering\includegraphics[height=3.25in]{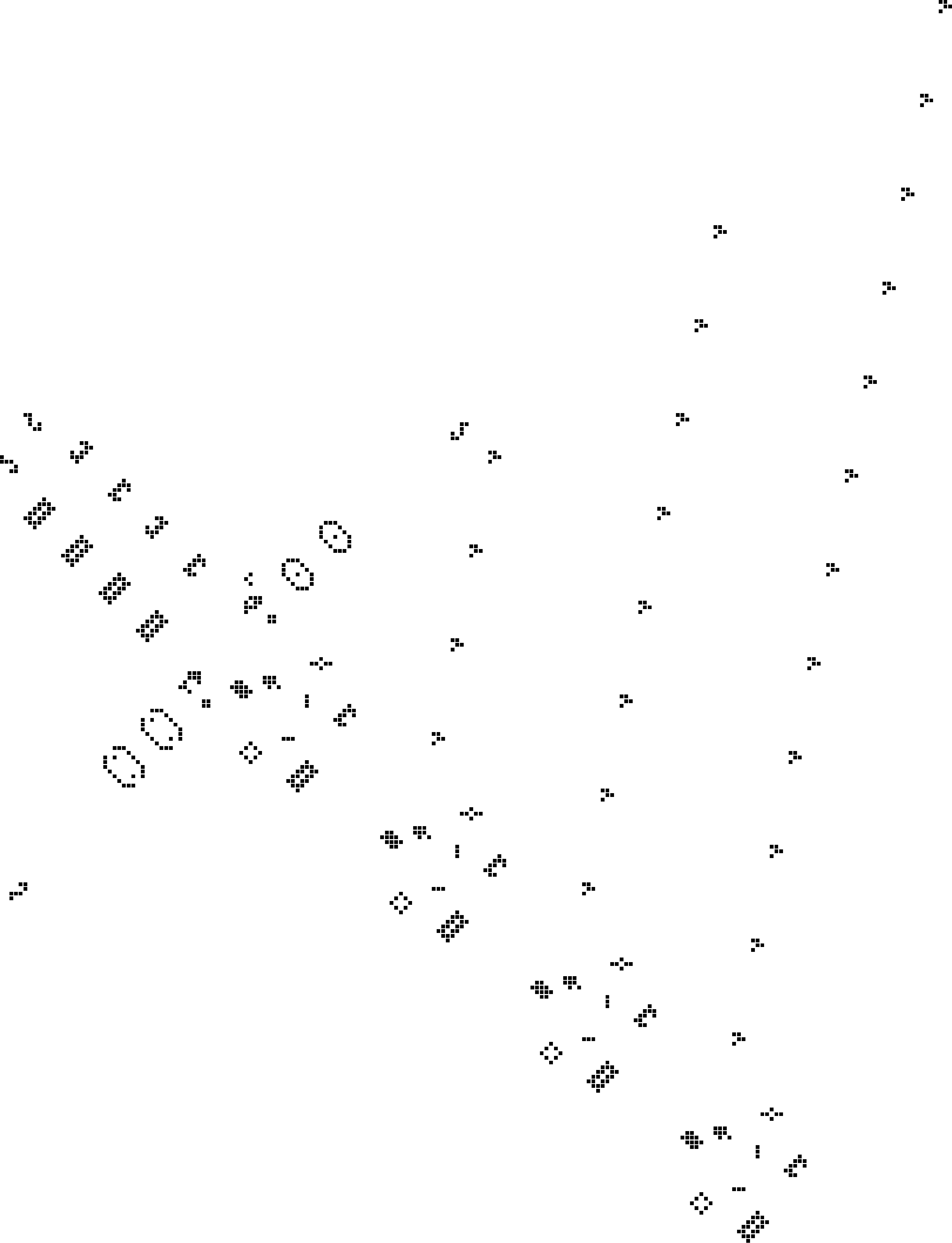}
\caption{A rake gun\index{rake gun} in HighLife.\index{HighLife} Four rows of replicators capped by eaters (left) interact to form pairs of bombers that move downwards and to the right. The sparks from each pair of bombers interact to form a trail of gliders that move upwards and to the right, filling a quarter of the plane with a quadratically growing\index{quadratic growth} number of live cells.}
\label{fig:b36s23-rakegun}
\end{figure}

It is our hypothesis that the rules most likely to support interesting patterns are the ones that are both fertile and mortal. Fig.~\ref{fig:map} depicts in rough terms this region of the rule space. In this section, we examine some specific rules where complex engineered patterns have already been discovered, in support of this hypothesis. Not coincidentally, these rules are both fertile and mortal, but they have differing Wolfram classes, demonstrating that Wolfram's classification does not accurately describe the existence of this sort of pattern.

\begin{description}
\descitem{HighLife (rule B36/S23)}\index{B36/S23}\index{HighLife} was investigated extensively by Bell\index{Bell, David I.}~\cite{Bel-94}. Many of its patterns and behaviors are similar to those in Life, because it differs from Life only in the comparatively rare case of a dead cell with six live neighbors. However, unlike Life, it features a small pattern known as a \emph{replicator}\index{replicator} that (in the absence of obstacles) replaces itself with two separated copies of the same pattern every twelve steps. Many nontrivial combinations of replicators with other patterns are known:
\begin{itemize}
\item Rows of replicators can be capped by oscillators or blocks, producing oscillators of arbitrarily large periods.
\item A single replicator together with a blinker oscillator produces the \emph{bomber},\index{bomber} a $c/6$ diagonal spaceship: when the replicator copies itself, one copy is shifted forward while the other copy destroys the blinker and replaces it with another blinker shifted forward by the same amount.
\item Combinations of bombers and replicators can produce puffers and rakes, moving objects that emit still lifes, gliders, or even rows of replicators.
\item Combinations of replicator-based oscillators can be used to make guns of arbitrarily high period that emit gliders, bombers, rakes, or other patterns (Fig.~\ref{fig:b36s23-rakegun}).
\item Dean Hickerson\index{Hickerson, Dean} has observed that a carefully timed sequence of replicators, interacting with a blinker, can be used to push it forwards by eight steps. Together with the bomber reaction that pulls a blinker towards a group of oscillators, it should be possible to use these reactions to construct very large spaceships that move at arbitrarily slow speeds.
\end{itemize}
As in Life, random fields in HighLife seem to eventually settle down to still lifes and small oscillators: the replicators cannot make progress through the other patterns that surround them. Thus, HighLife should probably be assigned the same Wolfram class as Life, either Class~II or Class~IV.

\begin{figure}[p]
\centering\includegraphics[width=4.25in]{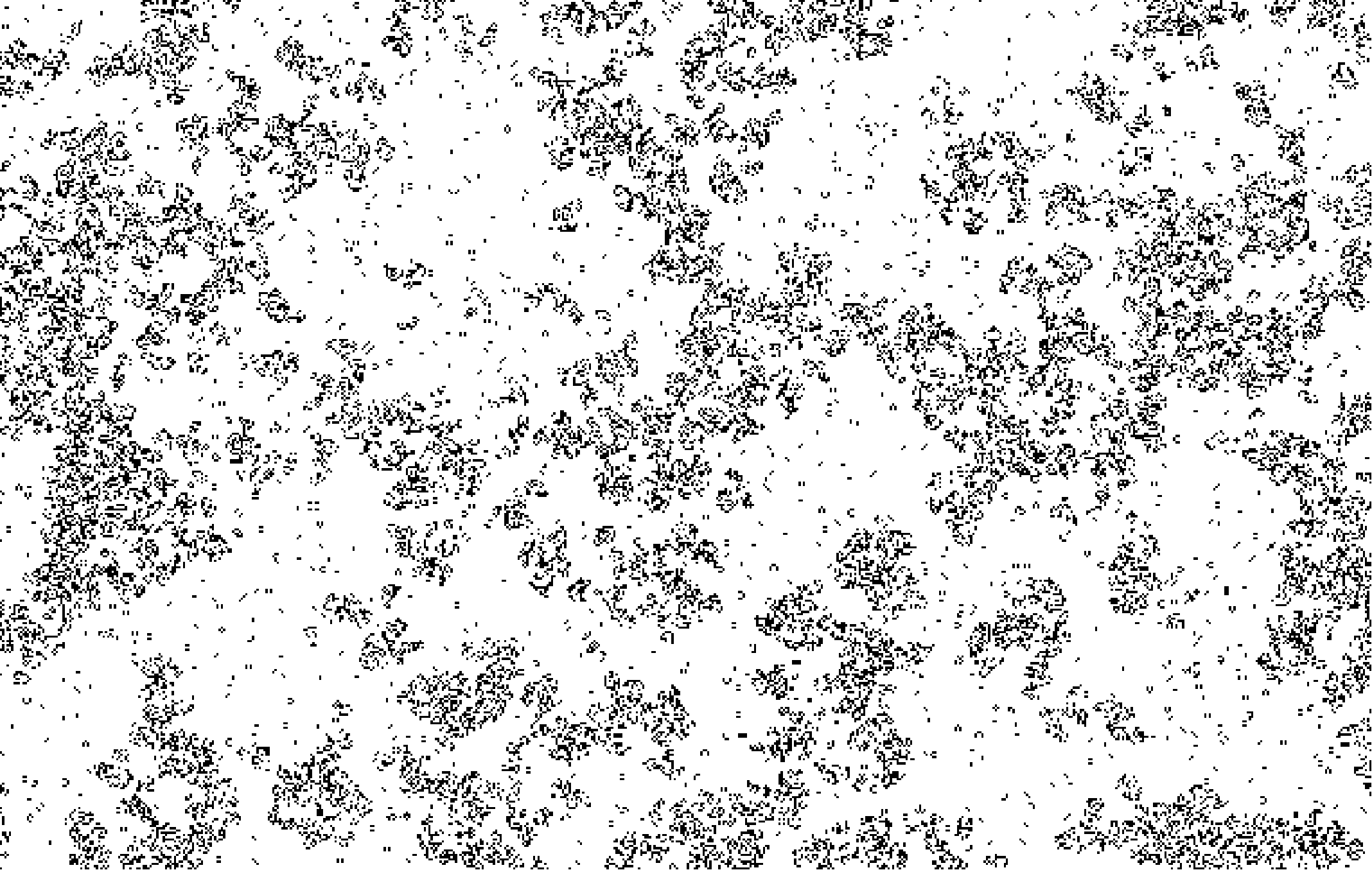}
\caption{B368/S12578,\index{B368/S12578} after 4000 steps from random initial conditions.}
\label{fig:b368s12578}
\end{figure}

\begin{figure}[p]
\centering\includegraphics[width=4in]{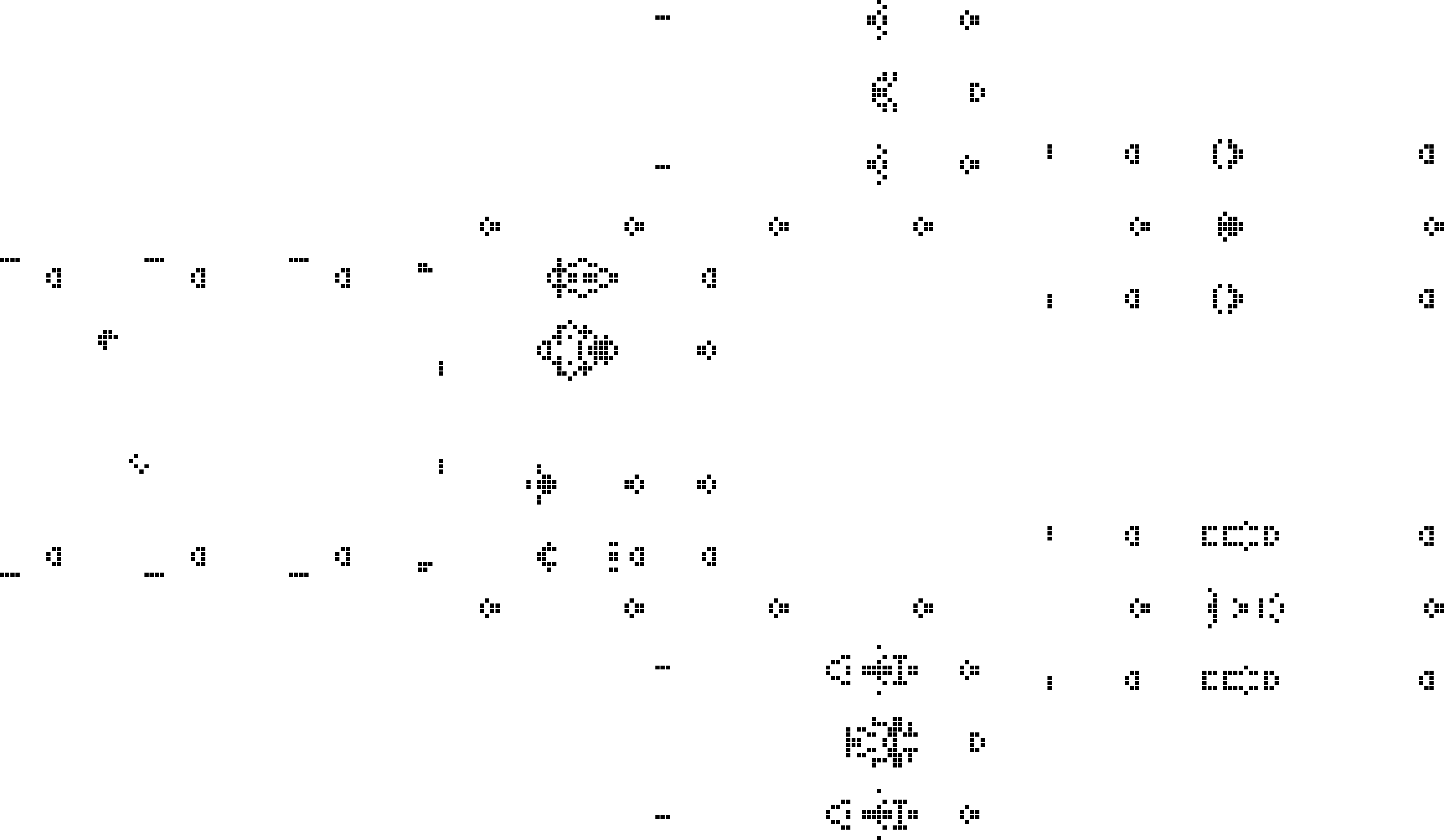}
\caption{Hickerson's\index{Hickerson, Dean} $28c/1200$ spaceship in B36/S245.\index{B36/S245}}
\label{fig:b36s245-28c1200}
\end{figure}

\begin{figure}[p]
\centering\includegraphics[width=4in]{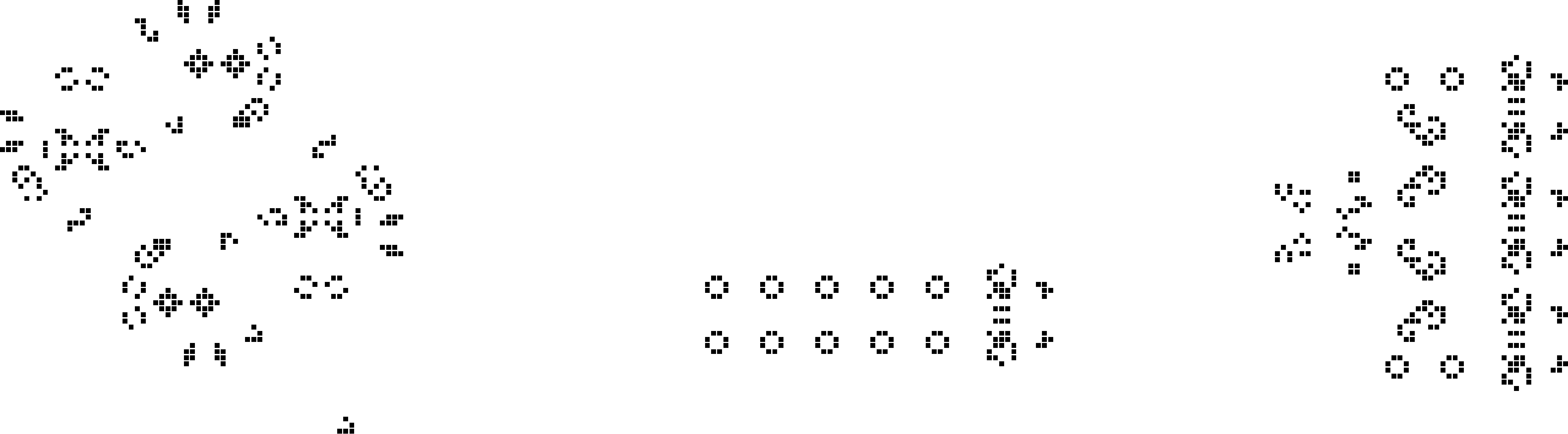}
\caption{A gun in B37/S23,\index{B37/S23} the $9c/28$ puffer formed from two R~pentominos,\index{R pentomino} and the $9c/28$ puddlejumper,\index{puddlejumper} a spaceship formed from five puffers.}
\label{fig:puddlejumper}
\end{figure}

\descitem{B368/S12578}\index{B368/S12578} like HighLife supports a simple replicator, in the form of a $1\times 5$ block of cells; it copies itself every 13 time steps. This replicator may be used to form oscillators of arbitrarily high periods; combinations of these oscillators can form guns for a small $c/8$ diagonal spaceship supported by this rule. As with HighLife, it seems likely that push-pull reactions can be used to form spaceships with arbitrarily slow speeds. However, with random initial conditions, this rule forms regions of chaotic activity interspersed with other regions of small oscillators and still lifes (Fig.~\ref{fig:b368s12578}), indicating a mixture of Class II and Class III behavior.

\descitem{B36/S245.}\index{B36/S245} Soon after the discovery of a replicator in HighLife, Mark Niemiec\index{Niemiec, Mark} found another rule with a replicator: B36/S245. The initial pattern for this replicator consists of a pair of ``shuttles'', sets of twelve live cells with a $3\times 6$ bounding box, in the shape of a capital letter D. Each shuttle, if sufficiently far from other patterns, repeats its shape in 102 generations, flipped 180 degrees, after laying a pair of ``eggs'' (period 4 oscillators). But, if an egg is already present, it produces a collection of sparks, which in the presence of the symmetrically placed shuttle end up hatching another replicator after 96 generations. The first few generations at which a copy of the replicator reappears are 102, 204, 300 (the first hatched egg), 306, 402, 504, 606, 702, and 708. Rule B36/S245 is also interesting because, like Life, it has many small spaceships, including a 3x3 period-7 diagonal glider, and a 4c/23 orthogonal spaceship. As in HighLife, rows of the replicators in this rule can be capped in various ways to form high-period oscillators and guns. Dean Hickerson also showed that the replicators in this rule can also be used to form high-period puffers that move 28 units of distance in either 600 or 1200 time steps. By combining groups of these puffers together, he found large spaceships that move at speed $14c/300$ and $28c/1200$ (Fig.~\ref{fig:b36s245-28c1200}). Like life, random states eventually but slowly settle down to scattered oscillators and still lifes, so it should be classified either as Class II or Class IV.

\descitem{Morley (B368/S245)}\index{B368/S245}\index{Morley} is named after Stephen Morley,\index{Morley, Stephen} who found many of its patterns; it has also been called Move. It contains small naturally-occurring spaceships of several periods, as well as a slow puffer with speed $13c/170$. Groups of puffers can be combined to form spaceships that move at the same speed, as well as breeders that emit streams of puffers moving sideways to the path of the breeder. This rule also supports several guns. Like life, random states eventually but slowly settle down to scattered oscillators and still lifes, so it should be classified either as Class II or Class IV.

\descitem{B37/S23}\index{B37/S23} is superficially similar to Life, but also supports a pattern unlike anything in Life in which a pair of R~pentominos\index{R pentomino} form a puffer that moves at speed $9c/28$, leaving a trail of pairs of pond still lifes behind them (Fig.~\ref{fig:puddlejumper},~center). Five of these puffers can be combined, with three in a front row and two in a back row, so that all of the ponds are destroyed by sparks from other puffers, resulting in a large $9c/28$ spaceship, the \emph{puddlejumper}\index{puddlejumper}~\cite{Epp-MSRI-02} (Fig.~\ref{fig:puddlejumper},~right). Along with its high-period spaceship, B37/S23 supports a glider gun found by Jason Summers\index{Summers, Jason} (Fig.~\ref{fig:puddlejumper},~left). Like B368/S12578, random initial conditions cause this rule to develop regions of chaotic activity interspersed with regions of scattered still life and oscillator patterns, indicating a mix of Class II and Class III activity.

\medskip
Several other rules also have large spaceships in which puffers interact to destroy all of the debris they would otherwise leave behind. These rules include B356/S23,\index{B356/S23} B356/S238,\index{B356/S238} B3678/S0345,\index{B3678/S0345} and B38/S02456\index{B38/S02456}~\cite{Fano}. In Life, Dean Hickerson's\index{Hickerson, Dean} $c/12$ diagonal Cordership,\index{Cordership} based on Charles Corderman's\index{Corderman, Charles} switch engine\index{switch engine} puffer, also has the same structure.

\begin{figure}[p]
\centering\includegraphics[width=3in]{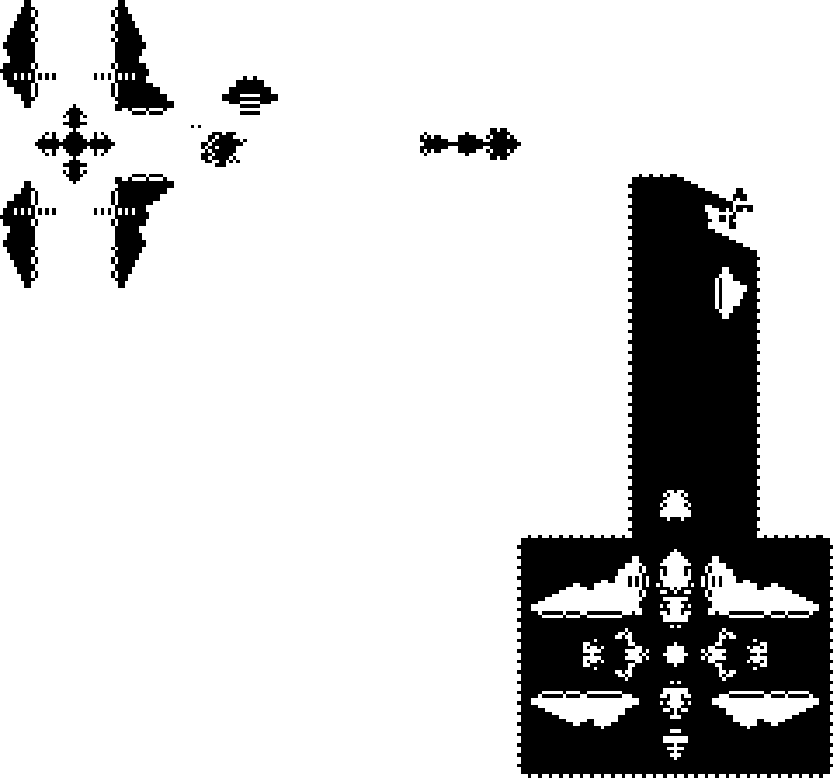}
\caption{A gun and an antigun interact in Day \& Night. Pattern by David Bell,\index{Bell, David} based on a reaction by Dean Hickerson,\index{Hickerson, Dean} from the Golly\index{Golly} pattern collection.}
\label{fig:dngag}
\end{figure}

\begin{figure}[p]
\centering\includegraphics[width=2.5in]{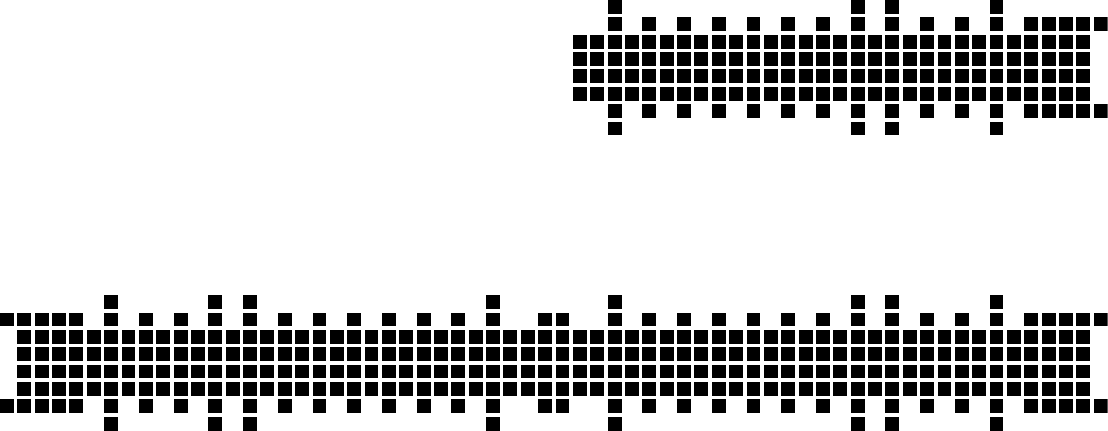}
\caption{A $c/7$ spaceship (top) and David Bell's spacefiller\index{spacefiller} (bottom) in Diamoeba\index{Diamoeba} (B35678/S5678).\index{B35678/S5678}}
\label{fig:diamoeba}
\end{figure}

\begin{figure}[p]
\centering\includegraphics[width=4.25in]{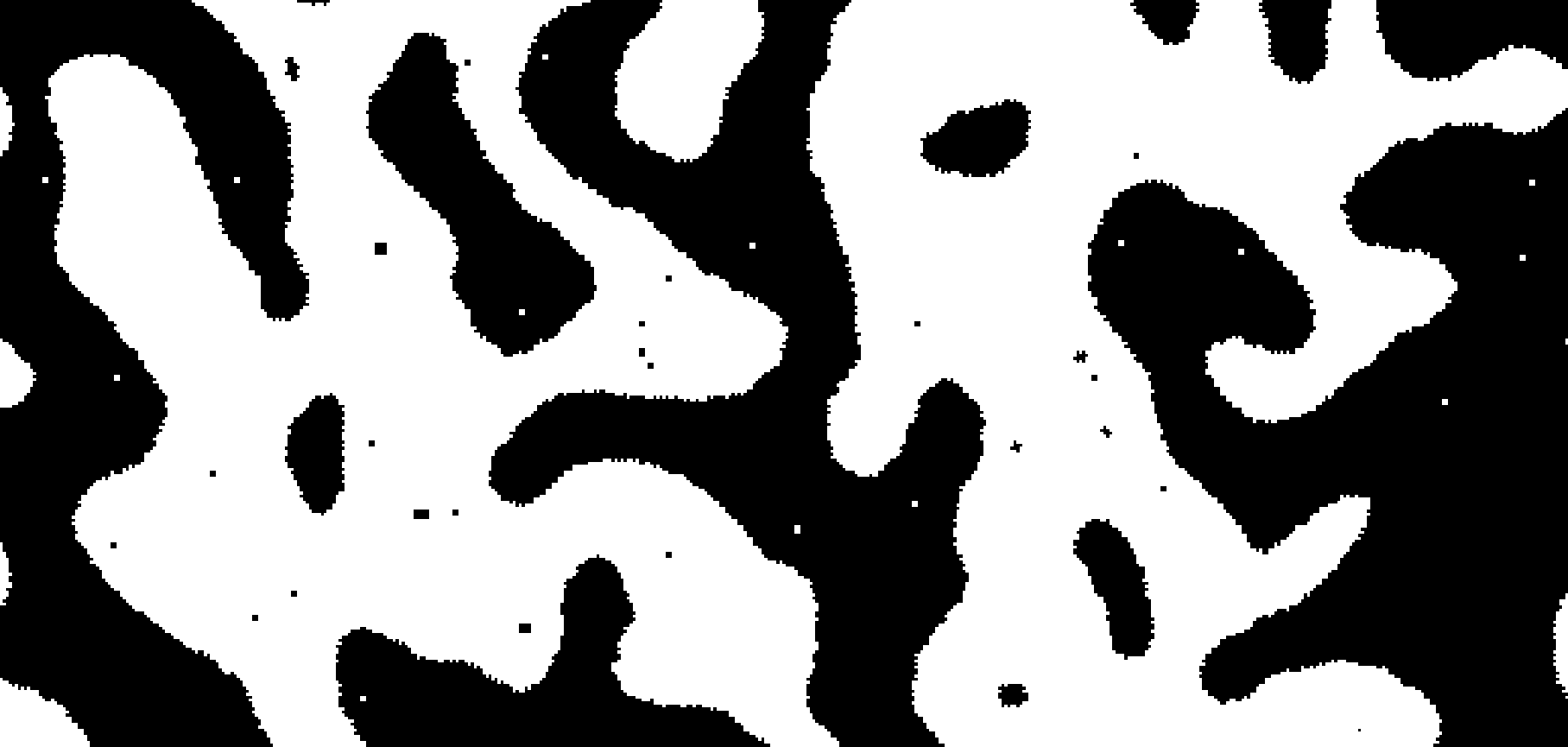}
\caption{After 500 steps starting from a random field in Anneal\index{Anneal} (B4678/S35678).\index{B4678/S35678}}
\label{fig:b4678s35678}
\end{figure}

\descitem{Day \& Night (B3678/S34678)}\index{B3678/S34678}\index{Day \& Night} is another rule investigated by Bell\index{Bell, David}~\cite{Bel-97}. It has the curious property that, if one exchanges the roles of live and dead cells, the rule's behavior is unchanged, because the numbers of dead neighbors that lead to a cell death are exactly the same as the numbers of live neighbors that lead to a birth, and the numbers of dead neighbors that lead to quiescence are exactly the same as the numbers of live neighbors that lead to survival. Day~\&~Night supports a small diagonal spaceship and many orthogonal ones, high-period oscillators, and patterns that mediate the boundaries between regions of live cells on a dead background and regions of dead cells on a live background (Fig.~\ref{fig:dngag}). When its initial state is a random field of cells with a probability $p < \frac12$ of being live, the live cells form clusters that gradually shrink, eventually forming a collection of scattered oscillators. When the probability of being live is greater than $\frac12$, the dead cells form shrinking clusters and one again gets a collection of scattered oscillators in the complementary dead-on-live world. When the probability of being live is exactly $\frac12$, the clusters and live and dead cells gradually grow and merge, so that the typical size of a cluster and the typical curvature of the cluster boundaries (ignoring small isolated oscillators) are monotonic functions of the number of steps of the rule; see Fig.~\ref{fig:b4678s35678} for an example of similar behavior in a different self-complementary rule. The boundaries between clusters are not a type of pattern that is directly addressed by Wolfram's classification, but if one interprets these boundaries as being largely stable or oscillatory, one can view the behavior of Day~\&~Night as belonging to Wolfram's Class~II.

\descitem{Diamoeba (B35678/S5678)}\index{B35678/S5678}\index{Diamoeba} was first investigated by Dean Hickerson.\index{Hickerson, Dean} In this rule, patterns tend to form large diamond shapes with an irregular and difficult-to-predict growth rate. It supports a small $c/7$ spaceship~\cite{Epp-MSRI-02} with some internal structure: between the head and the tail of the spaceship there is an extensible middle section, the boundary of which is lined by a sequence of one-cell and two-cell protrusions, and the time-space patterns formed by these protrusions simulate a simple one-dimensional cellular automaton. Hickerson offered a \$50 prize in 1993 for finding a quadratic growth pattern in 1998 Gravner and Griffeath~\cite{GraGri-AAM-98} asked more specifically whether there exists a pattern that eventually fills the entire plane with live cells. Both problems were solved in 1999 by David Bell,\index{Bell, David} whose solution combined two oppositely-oriented copies of the $c/7$ spaceship. Bell and Hickerson subsequently also found patterns for which the number of live cells grows linearly rather than quadratically, based on the same $c/7$ spaceship head. Under Wolfram's classification, Diamoeba appears either as Class I or Class II: random initial conditions lead to a state in which almost all cells are live, but in which there are very sparse clusters of oscillating dead cells.

\begin{figure}[p]
\centering\includegraphics[width=4.5in]{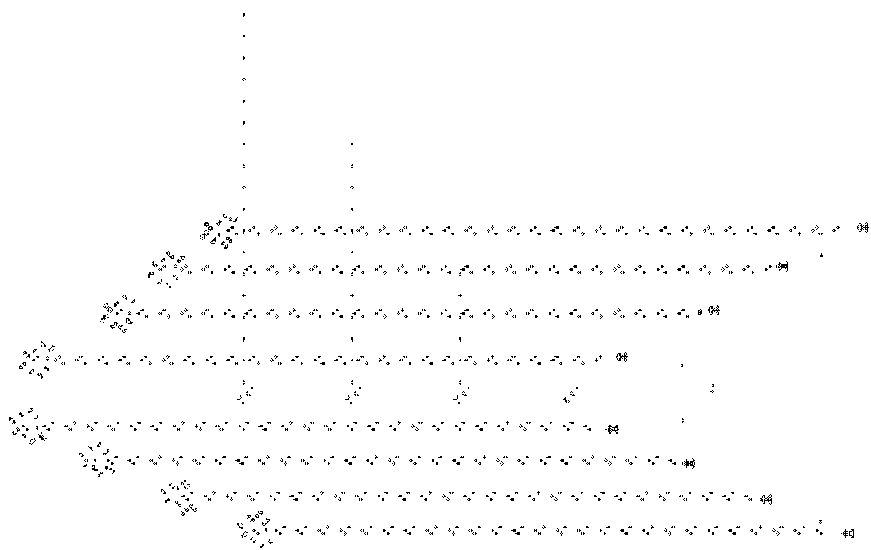}
\caption{A quadratic growth pattern in B35/S236\index{B35/S236}.}
\label{fig:b35s236-breeder}
\end{figure}

\begin{figure}[p]
\centering\includegraphics[height=2.15in]{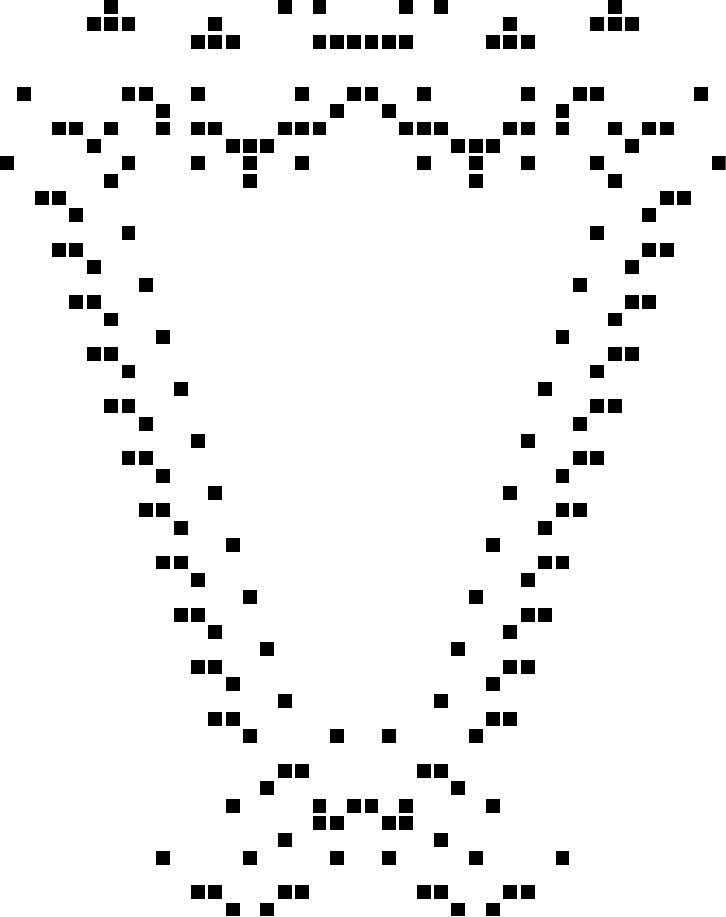}
\caption{A $c/2$ spaceship in B27/S0\index{B27/S0}.}
\label{fig:b27s0}
\end{figure}

\begin{figure}[p]
\centering\includegraphics{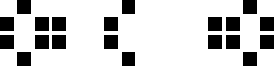}
\qquad\qquad\includegraphics{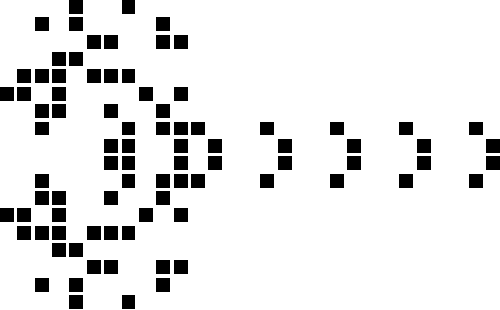}
\caption{Left: A spaceship sandwiched between two replicators in B25/S4\index{B25/S4} leads to pseudorandom behavior. Right: A glider gun in B24/S.\index{B24/S}}
\label{fig:2b2}
\end{figure}

\descitem{B35/S236}\index{B35/S236}~\cite{Epp-03} is near Life in rule space, but most patterns behave differently in the two rules. It has a small spaceship analogous to Life's glider, and as with the glider this spaceship frequently arises from small random seeds, but it moves at speed $2c/5$ orthogonally. This rule also supports small period-24 and period-68 oscillators; by combining multiple copies of either of these oscillators it is possible to form guns. Fig.~\ref{fig:b35s236-breeder} shows a pattern in which eight banks of three period-68 guns (with two additional period-68 spaceship reflectors per bank) send out streams of small $2c/5$ spaceships behind eight slower $c/3$ spaceships; when the $2c/5$ stream collides with the rear end of the $c/3$ spaceship, it sends a single spaceship towards the middle of the pattern. These spaceships collide to produce additional guns, leading to a pattern like Life's breeder in which the number of live cells grows quadratically with the number of steps. When started from a random initial condition, this rule remains chaotic, falling into Wolfram's Class~III.

\descitem{B27/S0}\index{B27/S0} supports a large $c/2$ spaceship (Fig.~\ref{fig:b27s0}) which, despite being found by an automated search~\cite{Epp-MSRI-02}, contains a large amount of structure similar to that of an engineered pattern. As oriented in the figure, it moves downwards; although the overall pattern has bilateral symmetry with even width, it has two head components that are themselves symmetric with odd width. Behind them two repeating patterns stretch out on either side, with 11 units of repetition per side, spreading out the pattern from the head components to a wider tail that includes two more of the same head components and six smaller equal tail components. However, when started in a random configuration, patterns in this rule remain highly and uniformly chaotic, placing it clearly in Class~III of Wolfram's classification.

\descitem{B25/S4} contains a small period-3 replicator,\index{replicator} and (like most B2 rules without B3/S1) supports small \emph{photons},\index{photon} spaceships that move at speed $c$. If a photon and a replicator are aligned on the same axis, their interaction will destroy that copy of the replicator and emit an oppositely-oriented photon. Placing a photon between two appropriately spaced replicators (Fig.~\ref{fig:2b2}, left) leads to a pattern in which the photon repeatedly bounces back and forth between copies of the replicators on both sides, following a pseudorandom walk in which the typical distance of the photon from its starting point at step~$n$ appears to be proportional to $\sqrt n$. After our 2001 discovery of this system, Tomas Rokicki\index{Rokicki, Tomas} found very efficient algorithms allowing its behavior to be simulated for billions of steps~\cite{Niv-07}. As Dean Hickerson\index{Hickerson, Dean} observed, a similar system can be set up in HighLife:\index{HighLife} the bomber reaction\index{bomber} allows two sets of replicators to play tug of war\index{tug of war} with a blinker.
Like most B2 rules, B25/S4 falls into Wolfram's Class~III.

\descitem{B24/S}\index{B24/S} supports a low-period photon gun\index{photon gun} (Fig.~\ref{fig:2b2}, right). Guns of this size and period are small enough to be found by automated search software; another, similar gun exists in B25/S45.\index{B25/S45} Again, these rules belong to Class~III.

\descitem{B2/S7}\index{B2/S7} was studied by Mart{\'\i}nez et al.~\cite{MarAdaMcI-08}.\index{Mart{\'\i}nez , Genaro Ju{\'a}rez}\index{Adamatzky, Andrew}\index{McIntosh, Harold V.} This rule supports spaceships, puffers, rakes, and a novel structure that Mart{\'\i}nez et al. termed \emph{avalanches}\index{avalanche} in which a diagonal chain of cells grows wider as it moves across the plane leaving a diamond-shaped trail of chaos behind it. They described methods of simulating Boolean circuits using spaceship collisions and implementing a finite-state memory with oscillators.
\end{description}

\noindent In contrast with these fertile and mortal rules, we describe a few rules that are infertile or immortal. Because of these properties, fewer structured patterns are known in these rules, but they may still exhibit other interesting behaviors.

\begin{description}
\descitem{Anneal (B4678/S35678)}\index{Anneal}\index{B4678/S35678} is mortal but infertile. Similarly to Day~\&~Night\index{Day \& Night}, any pattern in this rule has the same behavior if all live cells are replaced by dead cells and vice versa. And as with Day~\&~Night, its behavior on random initial configurations is to form growing clusters of live and dead cells, with scattered oscillators. The live cells predominate for initial configurations with probability greater than 0.5 of being live, the dead cells predominate when this probability is less than 0.5, and when the probability is exactly $\frac12$ (Fig.~\ref{fig:b4678s35678}) both types of clusters coexist with the cluster size and radius of curvature increasing over time. Thus, Wolfram's classification is incapable of distinguishing this rule from Day~\&~Night, although its nonrandom behavior is much more constrained.

\descitem{B1357/S1357}\index{B1357/S1357} is a fertile but immortal rule investigated by Edward Fredkin\index{Fredkin, Edward} in which every pattern is a replicator.\index{replicator} Cell $(x,y)$ is alive at step $i$ if and only if the number of ways that a chess king\index{chess king} could take $i$ steps to walk from an initially-live cell to $(x,y)$ is odd. When $i$ is divisible by a number $2^k$ that is larger than the size of the bounding box of the initial pattern $P$, the pattern at step $i$ consists of several disjoint copies of $P$, spaced at multiples of $2^k$ units apart, with the overall arrangement of these copies being identical to the arrangement of live cells that one would get at step $i/2^k$ starting from a single live cell.
Thus, although this rule supports replicators, which in many other rules lead to other sorts of complex behavior, in this rule there is nothing but replicators. If a starting state has all cells set to live or dead uniformly and independently at random, then the same is true at each subsequent step, so this rule displays no structure whatsoever when run under random initial conditions and falls into Wolfram's Class~III.

\begin{figure}[t]
\centering\includegraphics[height=2.5in]{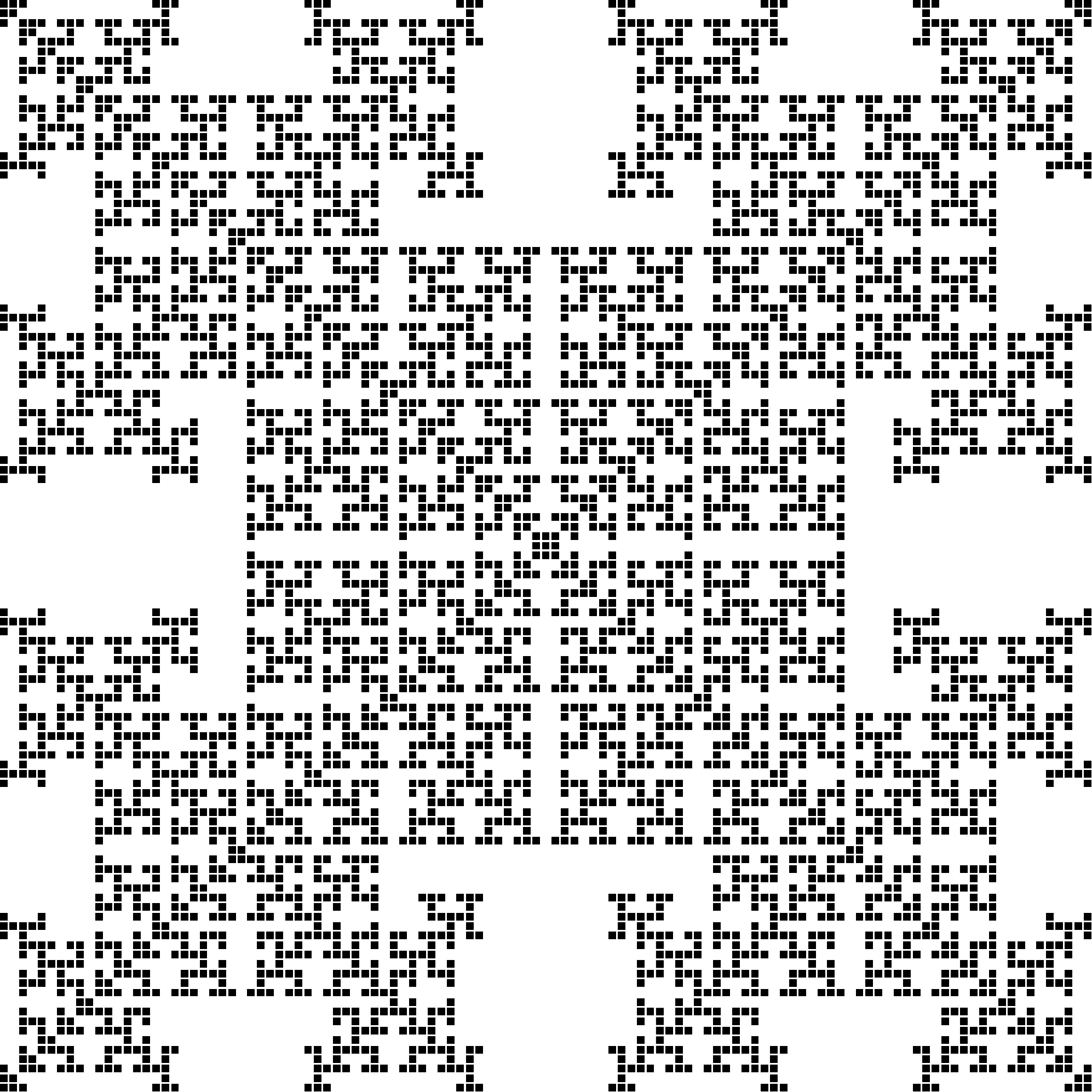}
\caption{The space-filling tree structure formed by B1/S012345678\index{B1/S012345678} after 57 steps starting from a single live cell.}
\label{fig:H-tree}
\end{figure}

\descitem{B1/S012345678}\index{B1/S012345678} is another fertile and immortal rule, introduced as a model of snowflake formation by Packard~\cite{Pac-84} and one of several Life-like rules later mentioned by Wolfram and Packard~\cite{WolPac-JSP-85},\index{Wolfram, Stephen}\index{Packard, N. H.}.
In this rule a starting state consisting of a single live cell leads to a pattern that fills the plane with a fractal tree structure---see Fig.~\ref{fig:H-tree}. The ratio of live cells to dead cells in the eventual stable pattern is exactly $\frac49$~\cite{GraGri-AAM-98}; however, Dean Hickerson\index{Hickerson, Dean} has found other starting patterns for this rule that lead to different densities~\cite{GraGri-NL-09}. See~\cite{GraGri-NL-09} for similar results in all the related \emph{Packard snowflake}\index{Packard snowflake} Life-like rules with rule strings of the form B1\textit{xxx}/S012345678.  When run from a random initial state, B1/S012345678 very quickly stabilizes, putting it in Class~II.

\begin{figure}[t]
\centering\includegraphics[width=4in]{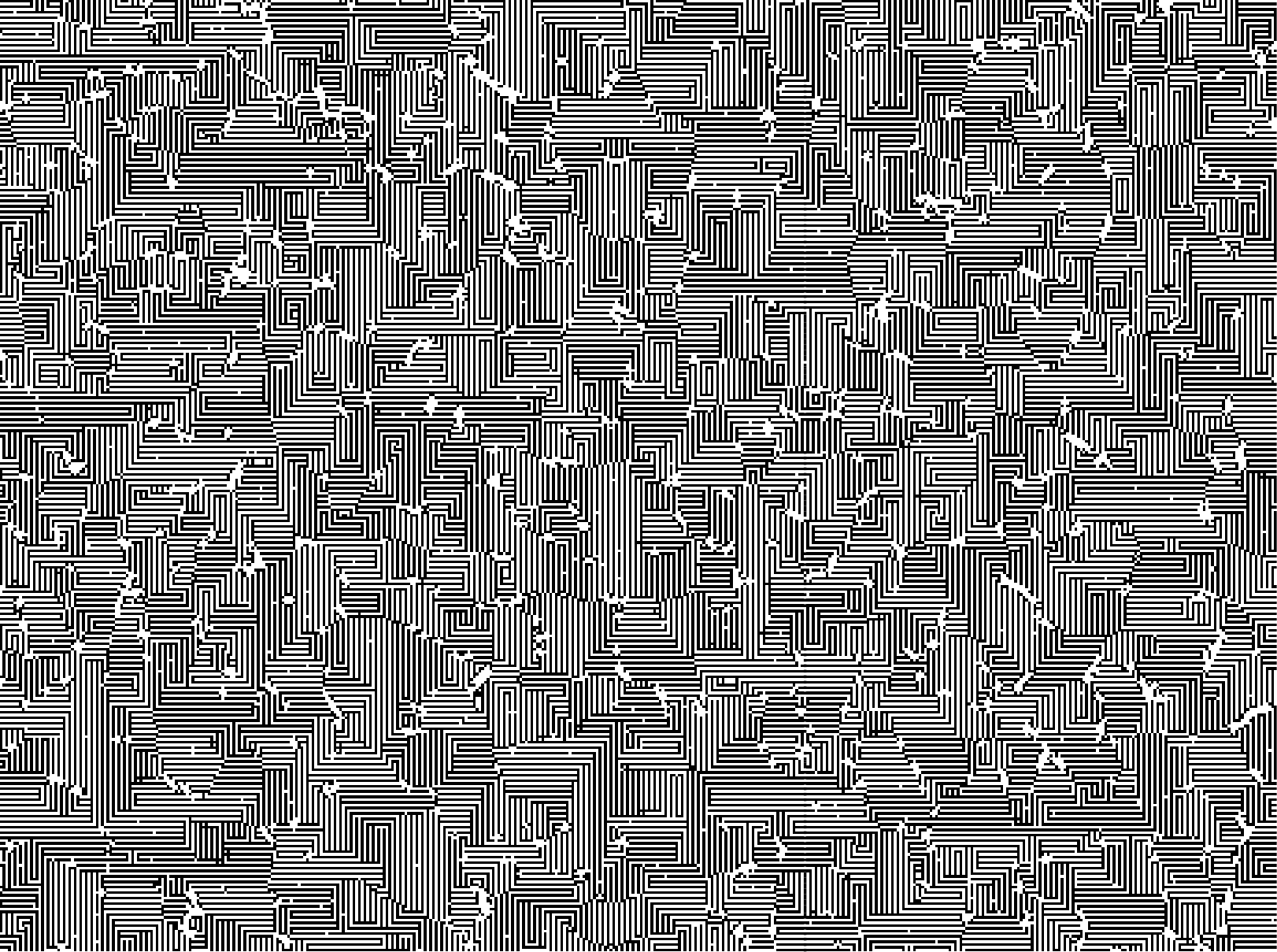}
\caption{Regions of horizontal and vertical stripes created by B4/S01234.}
\label{fig:b4s01234}
\end{figure}

\descitem{B4/S01234}\index{B4/S01234} is both infertile and immortal. When run from a random initial state it crystallizes\index{crystallization} into regions filled with horizontal stripes of live and dead cells, mixed with similar regions with vertical stripes (Fig.~\ref{fig:b4s01234}); some cells on the borders between regions oscillate with low periods. Similar striped patterns also form in B35/S234578\index{B35/S234578}, a mortal rule that seems extremely likely to be fertile: although we do not know of a spaceship nor a proof that any other pattern is a growth pattern, most finite starting patterns in B35/S234578 form round regions with a linearly growing radius, within which these same patches of horizontal and vertical stripes predominate.

\begin{figure}[t]
\centering\includegraphics[height=3in]{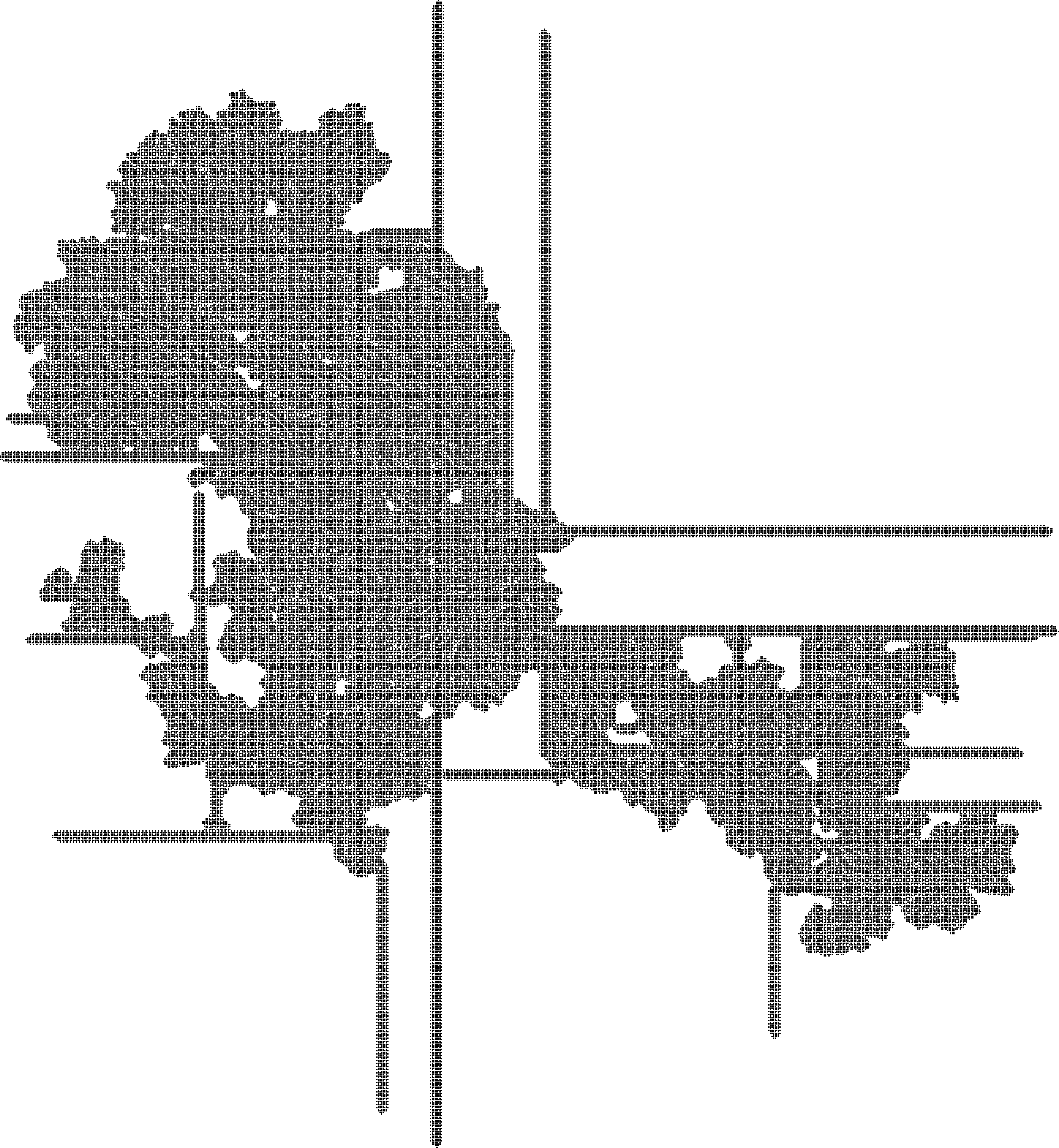}
\caption{A pattern resulting from approximately 1000 steps of Life Without Death\index{Life Without Death} starting from a small random seed. The long straight patterns extending from the central mass are ladders.}
\label{fig:lwod}
\end{figure}

\descitem{Life Without Death (B3/S012345678)}\index{B3/S012345678}\index{Life Without Death} (also known as Inkspot~\cite{TofMar-87}) has the same birth rules as Life, but disallows any death of a live cell, causing it to be immortal. It is fertile, despite having no spaceships: its patterns frequently develop \emph{ladders}\index{ladder} consisting of a growing tip that moves in a straight line leaving an immortal trail of live cells behind it~(Fig.~\ref{fig:lwod}). Ladders can be arranged to simulate any Boolean circuit, showing that it is P-complete to determine the value of a cell at a future state of the automaton~\cite{GriMoo-CS-96}; this implies that it is unlikely for there to exist an algorithm that solves this cell value problem significantly more quickly than the naive algorithm that simply simulates the rule for the desired number of steps. In some sense this circuit simulation result tests the limits of our classification by showing that complex engineered patterns with a modular structure may exist even in immortal rules.
\end{description}

\section{Beyond growth and decay}

Our definitions of growth and decay make the assumption that the patterns to be analyzed have finitely many live cells, set on a background of quiescent dead cells. However, even for the restricted family of two-dimensional semitotalistic automata that we study here, these are not the only possibilities. Many automata, even those that may be wildly chaotic when started randomly, may support background patterns that are periodic in both time and space, as well as finite perturbations to these periodic background patterns that move and interact similarly to the way Life's gliders move and interact.

As one particular case of a periodic background, we have made some preliminary investigations of rules in which a birth occurs with zero live neighbors and a death occurs with eight live neighbors. In these rules, if a pattern starts with finitely many live cells on a background of dead cells, it will continue to have finitely many live cells on a background of dead cells in every even step, but in the odd steps the pattern is reversed: there are finitely many dead cells on a background of infinitely many live cells. A few of these rules have already shown interesting behavior:

\begin{figure}[t]
\centering\includegraphics[width=4in]{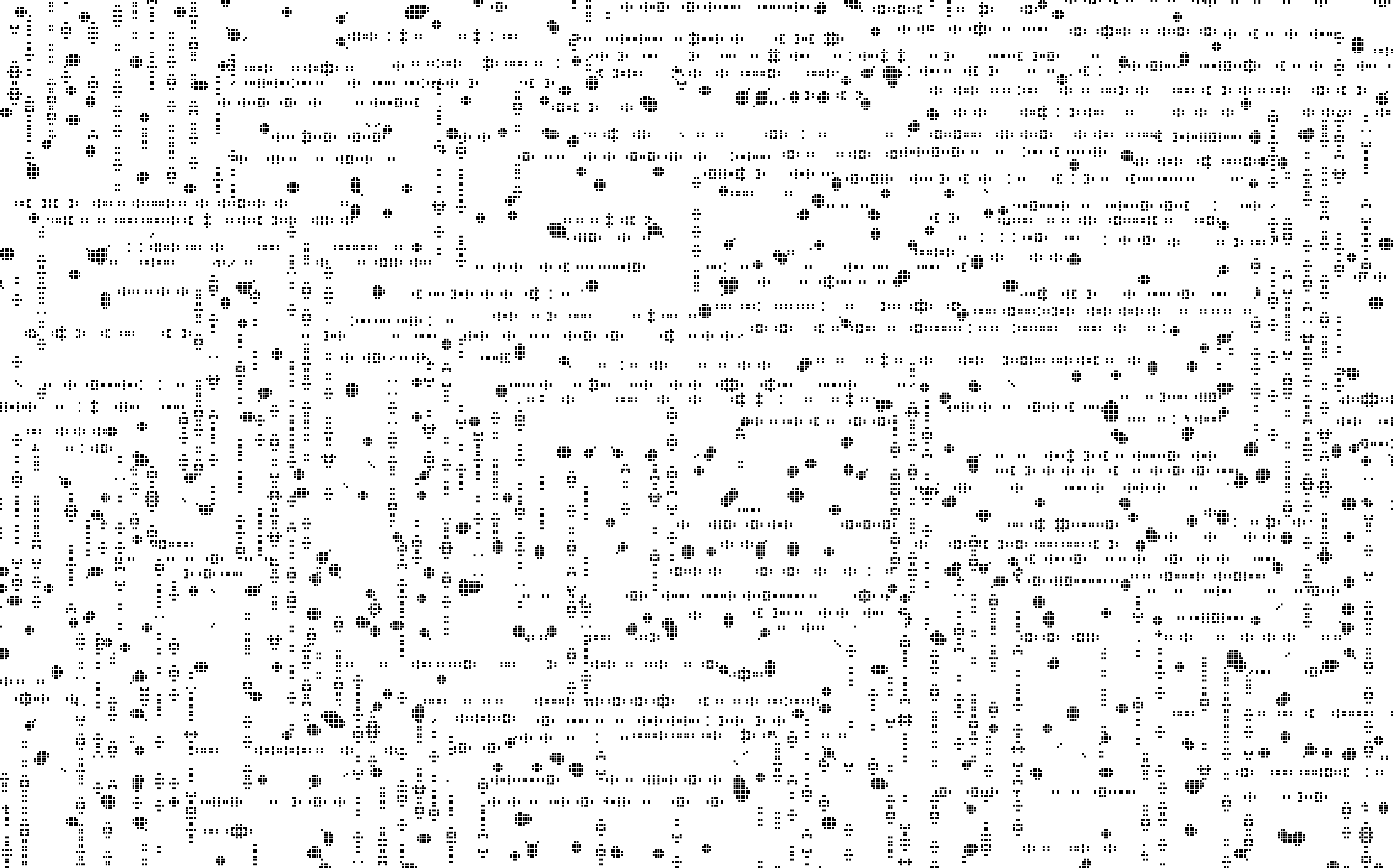}
\caption{Replicator chaos in B017/S1.}
\label{fig:b017s1}
\end{figure}

\begin{description}
\descitem{B01245/S0125}\index{B01245/S0125} is a strobing\index{strobing} version of Day~\&~Night\index{Day \& Night}. Any pattern in Day~\&~Night will behave identically in this rule, except that in odd steps the live and dead cells will reverse roles. Similar strobing versions exist for any rule that is symmetric under live-dead reversal, such as Anneal.\index{Anneal}

\descitem{B017/S1}\index{B017/S1} supports two different replicators, with periods 8 and 14. Initial states in which cells are set to be alive independently at random with either a low or high density of live cells eventually become dominated by oscillators formed by rows of these replicators, bounded at each end by stable blobs of cells (Fig.~\ref{fig:b017s1}); however, for intermediate densities of initial live cells (around 25\%), the pattern instead coagulates into larger stable blobs of cells that are all alive or all dead in alternating phases.

\descitem{B013468/S02}\index{B013468/S02} (a rule investigated by the author in 2002) has a small spaceship, and a very small period-36 double-barreled gun that in some phases fits into a $4\times 6$ bounding box. As in Day~\&~Night\index{Day \& Night}, it also has several larger high-period $c/2$ puffers and spaceships, as well as ladders\index{ladder} like those in Life Without Death. Fig.~\ref{fig:b013468s02} shows a pattern formed from eight puffers that would, if by themselves, leave trails of domino oscillators. Pairs of puffers combine to form high-period spaceships with large trailing sparks, one pair of spaceships combines to form a much messier puffer, and the other pair of spaceships combine to form a rake, whose output spaceships crash into the puffer trail to eliminate some unwanted debris, leaving behind a sequence of ladders with a quadratic growth rate in the number of non-background cells. Random initial conditions tend to settle down to large regions of still life, oscillator, and spaceship patterns, like those in Life, but as in Day~\&~Night\index{Day \& Night} some of these regions have live cells on dead backgrounds in even steps and dead cells on live backgrounds in odd steps, while for other regions this pattern is reversed. The boundaries of these regions are chaotic and send spaceships into the interiors of the regions, so patterns tend to stabilize much more slowly than in Life. As in Day \& Night, sparse random conditions eventually settle down to live cells on dead backgrounds on even steps and dense random conditions eventually settle down to the opposite parity, with the transition between these two phases occurring at roughly a 46\% ratio of live to total cells.

\begin{figure}[t]
\centering\includegraphics[width=3in]{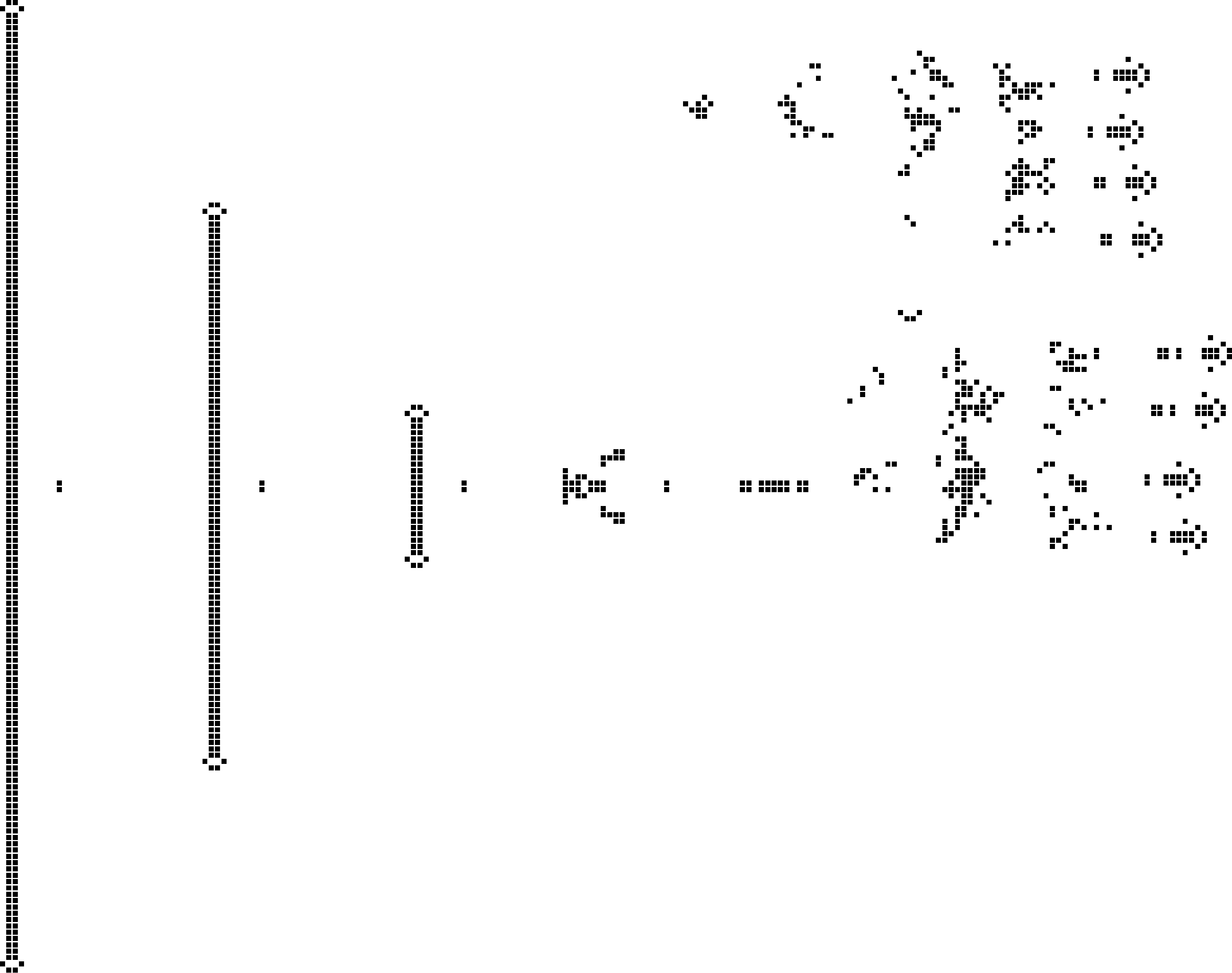}
\caption{A quadratic growth pattern in B013468/S02.}
\label{fig:b013468s02}
\end{figure}

\descitem{B01367/S0124}\index{B01367/S0124} has an unusual replicator in the shape of a W~pentomino\index{W pentomino} with the property that the two copies formed from this replicator are turned at right angles from its original orientation.

\descitem{B01367/S012}\index{B01367/S012} has replicators of periods 20 and 22 that can interact to form spaceship guns.

\descitem{B02346/S023}\index{B02346/S023} has very tiny replicators consisting of two live cells a knight's\index{knight's move} move apart; the copies this pattern makes of itself spread themselves out along a line of slope $\pm 2$ or $\pm 1/2$.
\end{description}

\noindent
One may generalize our definitions of fertility and mortality for such rules by replacing the finite sets of live cells in the definition by a finite set of cells that is different from the background: a fertile combination of a rule and a background is one for which some finite perturbation to the background escapes any bounding box, and a mortal combination of rule and background is one in which some finite perturbation to the background eventually stabilizes so that only the background remains. We know little about which combinations of rules and backgrounds are likely to be both fertile and mortal, but such knowledge would be very helpful as a guide in exploring the limitless possibilities these combinations have to offer.

\section*{Acknowledgements}

We thank Calcyman, David Griffeath, Dean Hickerson, Nathaniel Johnston, Harold McIntosh, and Andrew Trevorrow for helpful comments on a draft of this chapter. The author is supported in part by NSF grant 0830403 and by the Office of Naval Research under grant N00014-08-1-1015.

{\raggedright
\bibliographystyle{abuser}
\bibliography{lifelike}}

%\printindex
\end{document}